\documentclass[longauth]{aa} 

\usepackage{graphicx}
\usepackage{txfonts}
\usepackage{hyperref}
\defcitealias{alo15}{Paper~I}
\defcitealias{alo21}{Paper~II}

\begin{document}

\title{Variable stars in the VVV globular clusters}

\subtitle{III. RR~Lyrae stars in the inner Galactic globular clusters}

  \author{Javier Alonso-Garc\'{i}a \inst{1,2}
  \and
  Leigh C. Smith\inst{3}
  \and
  Jason L. Sanders\inst{4}
  \and
  Dante Minniti\inst{5,6}
  \and
  M\'{a}rcio Catelan\inst{7,2}
  \and
  Gonzalo Aravena Rojas\inst{8,9}
  \and
  Julio A. Carballo-Bello\inst{10}
  \and
  Jos\'e G. Fern\'andez-Trincado\inst{11}
  \and
  Carlos E. Ferreira Lopes\inst{12,2}
  \and
  Elisa R. Garro\inst{13}
  \and
  Zhen Guo\inst{8,2,14}
  \and
  Maren Hempel\inst{5,15}
  \and
  Philip W. Lucas\inst{14}
  \and
  Daniel Majaess\inst{16}
  \and
  Roberto K. Saito\inst{17}
  \and
  A. Katherina Vivas\inst{18}
  }
  \institute{Centro de Astronom\'{i}a (CITEVA), Universidad de Antofagasta,
  Av. Angamos 601, Antofagasta, Chile\\
  \email{javier.alonso@uantof.cl}
  \and
  Millennium Institute of Astrophysics, Nuncio Monse\~nor Sotero Sanz 100,
  Of. 104, Providencia, Santiago, Chile
  \and
  Institute of Astronomy, University of Cambridge,
  Madingley Road, Cambridge, CB3 0HA, UK
  \and
  Department of Physics and Astronomy, University College London,
  London WC1E 6BT, UK
  \and  
  Instituto de Astrof\'isica, Facultad de Ciencias Exactas,
  Universidad Andr\'es Bello, Fern\'andez Concha 700,
  Las Condes, Santiago, Chile
  \and
  Vatican Observatory, Vatican City State V-00120, Italy
  \and
  Instituto de Astrof\'{i}sica, Facultad de F\'isica,
  Pontificia Universidad Cat\'{o}lica de Chile,
  Av. Vicu\~na Mackenna 4860, 7820436 Macul, Santiago, Chile
  \and
  Instituto de F\'isica y Astronom\'ia, Universidad de Valpara\'iso, Av. Gran Breta\~na 1111, Playa Ancha, Casilla 5030, Chile
  \and
  Vera C. Rubin Observatory, AURA/NOIRLab, Chile
  \and
  Instituto de Alta Investigaci\'on, Universidad de Tarapac\'a, Casilla 7D, Arica, Chile
  \and
  Instituto de Astronom\'ia, Universidad Cat\'olica del Norte, Av. Angamos 0610, Antofagasta, Chile
  \and
  Instituto de Astronom\'ia y Ciencias Planetarias, Universidad de Atacama, Copayapu 485, Copiap\'o, Chile
  \and
  ESO - European Southern Observatory, Alonso de C\'ordova 3107, Vitacura, Santiago, Chile
  \and
  Centre for Astrophysics, University of Hertfordshire, Hatfield AL10 9AB, UK
  \and
  Max Planck Institute for Astronomy, K\"onigstuhl 17, 69117 Heidelberg, Germany
  \and  
  Mount Saint Vincent University, Halifax, Canada
  \and
  Departamento de F\'{i}sica, Universidade Federal de Santa Catarina, Trindade 88040-900, Florian\'{o}polis, SC, Brazil
  \and
  Cerro Tololo Inter-American Observatory/NSF's NOIRLab, Casilla 603, La Serena, Chile
  }

  \date{ }
   

   \abstract
   {High reddening near the Galactic plane hampers observations and
     proper characterization of the globular clusters (GCs) located
     toward the inner regions of the Milky Way.}
   {The VISTA Variables in the V\'ia L\'actea (VVV) survey observed the
     Galactic bulge and adjacent disk for several years, providing
     multi-epoch, near-infrared images for 41 Galactic GCs. Detecting
     RR~Lyrae variable stars belonging to these GCs will aid in their
     accurate parameterization.}
   {By fully leveraging the astrometric, photometric, and variability
     VVV catalogs, we searched for RR~Lyrae stars associated
     with GCs. Our selection criteria, based on proper motions,
     proximity to the cluster centers, and distances inferred from
     their period-luminosity-metallicity relations, enable us to
     accurately identify the RR~Lyrae population in these GCs and
     determine color excesses and distances to these poorly studied GCs in
     a homogeneous manner. Since the VVV catalogs cover from the
     innermost regions of the GCs to their outskirts, we can provide a
     comprehensive picture of the entire RR~Lyrae population in these
     GCs.}
   {We have discovered significant RR~Lyrae populations in two highly
     reddened Galactic GCs: UKS~1 and VVV-CL160, previously unknown to
     host RR~Lyrae stars. Additionally, we have detected one RR~Lyrae
     candidate in each of Terzan~4 and Terzan~9, also new to RR~Lyrae
     detection. We further confirm and increase the number of RR~Lyrae
     stars detected in 22 other low-latitude Galactic GCs.  The
     RR~Lyrae distances place most of these GCs within the Galactic
     bulge, aligning well with the few GCs in our sample with reliable
     {\em Gaia} or {\em Hubble} Space Telescope measurements. However,
     most of the VVV GCs lack accurate {\em Gaia} distances, and
     literature distances are generally significantly smaller than
     those derived in this work. As a byproduct of our analysis, we
     have obtained the proper motions for all the VVV GCs,
     independently confirming {\em Gaia} results, except for two of
     the most reddened GCs: UKS~1 and 2MASS-GC02.}
   {}


   \maketitle

%

\section{Introduction}
\label{sec_intro}

The Galactic globular clusters (GCs) constitute some of the most
massive and outstanding stellar aggregates in the Milky Way. Their
location in the Milky Way allows us to explore in detail their stellar
populations, as we can separate their individual stars down to a limit
not available for their counterparts in other galaxies. Recent
photometric and spectroscopic studies of their member stars from the
ground and from space \citep[e.g.,][]{mil17,mar19,bau23} have led to a
deeper understanding of the nature of these objects. The advent of the
{\em Gaia} mission \citep{gai16} has been a crucial landmark, not only
in more accurately establishing the physical parameters of the
Galactic GCs \citep{bau19,vas21}, but also for better understanding
their origin \citep{mas19,bel24}.

However, the GCs located closer to the Galactic plane suffer the
effects of high extinction, which prevents their proper
characterization when we use optical wavelengths. Near-infrared
observations are better suited, as extinction is highly diminished at
these wavelengths. Nevertheless, other additional problems present in
these regions, such as the elevated density of field stars, further
hamper the study of these GCs. In order to improve our analyses, we
need to complement the usual photometric studies of the
color-magnitude diagram (CMD) with other tools.

Paramount among them is the search for intrinsically bright pulsating
stars such as RR~Lyrae variables, which not only provide hints about the old
ages of the GCs but can also allow us to precisely measure
observational parameters such as distances and reddenings \citep{cat15}.
Near-infrared time-series observations allow us not only to more easily detect
these stars despite the high extinction of these low-latitude regions
but to more precisely provide the parameters of the GCs by using the
tighter period-luminosity-metallicity (PLZ) relations at these
wavelengths \citep[e.g.,][]{cat04,bra19a,cus21}.

The VISTA Variables in the V\'ia L\'actea survey \citep[VVV;][]{min10}
provides multi-epoch near-infrared observations of the Galactic bulge
and disk, essential for identifying and characterizing the variable
stellar populations of the GCs located toward these regions.
Approximately one-quarter of the known GCs in the Milky Way are
located inside the footprint of the VVV survey. In \citet{alo15} and
\citet{alo21}, hereafter referred to as \citetalias{alo15} and
\citetalias{alo21}, respectively, we searched for variable stars in
seven low-latitude GCs with different metallicities, presenting our
detection algorithm and refining the parametrization of the PLZ
relations for the near-infrared filters in the VVV survey. In the
current paper, the third in the series, we aim to take full advantage
of recent photometric catalogs produced with the VVV data to further
investigate the RR~Lyrae pulsating stellar populations in the
low-latitude GCs within the VVV footprint and use them to better
characterize the entire sample of these GCs.

\section{Observations and datasets}
\label{sec_obs}
The VVV survey \citep{min10,sai12b} used the 4.1m VISTA telescope at
Cerro Paranal Observatory to obtain photometric observations of the
Galactic bulge ($-10\fdg0 \le l \le +10\fdg4$, $-10\fdg3 \le b \le
+5\fdg1$) and an adjacent disk area ($294\fdg7 \le l \le 350\fdg0$,
$-2\fdg25 \le b \le +2\fdg25$) of the Milky Way in the $Z$, $Y$, $J$,
$H$, and $K_s$ near-infrared filters. The camera on board the VISTA
telescope provided $1.5\times1.1$ deg$^2$ wide-field images with a
resolution of 0.34\arcsec per pixel. The $K_s$ multi-epoch
observations from the VVV survey allow for the inspection of
photometric variability among the detected sources.

As we detail in the following sections, we use the VVV Infrared
Astrometric Catalogue 2 (VIRAC2) and VIrac VAriable Classification
Ensemble (VIVACE) databases to carry out such a variability search and
characterization in the VVV GCs. The VIRAC2 catalog \citep{smi25}, an
updated version of the VIRAC database \citep{smi18} now built using
point spread function (PSF) photometry, provides positions, proper
motions, mean magnitudes, and light curves for all detections from the
VVV observations. VIRAC2 uses the DoPHOT software \citep{sch93,alo12}
to detect sources and extract their PSF photometry from the VVV images
\citep{alo18} in the different filters and epochs. VIRAC2 extends the
timeline of the original VVV observations by incorporating the
additional $J$, $H$ and $K_s$ images taken of these regions in the
more recent VVV extended survey \citep[VVVX,][]{min18d,sai24}. Each
VVV field used in this work was observed at least twice in $Z$, $Y$,
$J$, and $H$, and at least 70 times in the $K_s$ filter. Photometry
from the different sources is later cross-matched and calibrated into
the VISTA photometric system \citep{gon18} to produce the multi-epoch
near-infrared light curves that were used by VIVACE for detecting
variable sources. The VIVACE database \citep{mol22} is a catalog of
variable sources which, using machine-learning techniques, identifies,
classifies, and parameterizes variable stars in the VVV original
footprint. It uses the VIRAC2 catalog and light curves as input for
its variability detections, yielding a catalog of $\sim$1.4 million
variable stars, for which it provides, among other parameters,
position, variable types, periods, amplitudes, and mean magnitudes in
the different VVV near-infrared filters.

\section{Analysis}
\label{sec_analysis}
The VIVACE catalog already provides a list of variable sources in the
region of the VVV GCs. A comparison with the seven GCs sampled in
\citetalias{alo21} reveals that nearly all of the RR~Lyrae stars
reported there are also present in the VIVACE catalog. In addition to
other light curve parameters, VIVACE provides precise periods for the
RR~Lyrae stars, enabling the construction of phased light curves from
our VIRAC2 photometry, of the same quality as those presented in
\citetalias{alo21}. VIVACE also offers sub-classifications for the RR
Lyrae stars, distinguishing between fundamental-mode (RRab) and
first-overtone (RRc) types. Thus, we decided not to further search for
more RR~Lyrae variable stars in the remaining VVV GCs and instead to
inspect all the VIVACE detections in their surroundings.

\begin{figure*}
  \centering
  \includegraphics[scale=1.0]{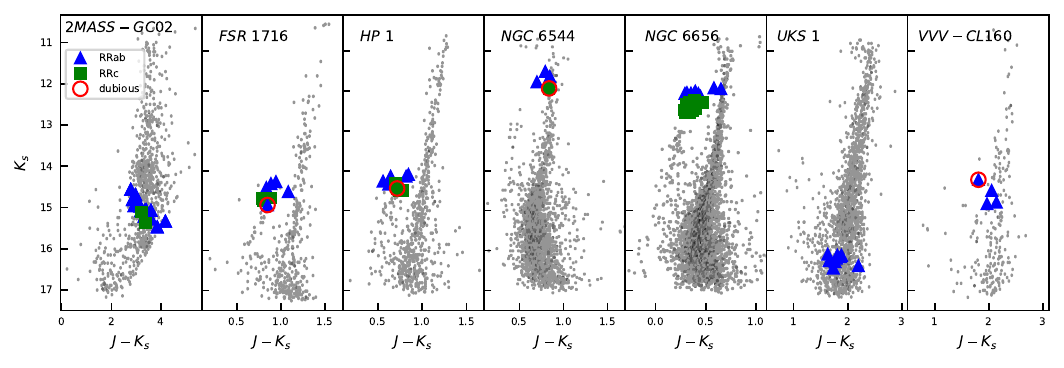}
  \caption{Near-infrared CMDs for selected VVV GCs with assigned
    RR~Lyrae members. The diagrams include only stars located in the
    innermost GC regions ($r\leq1.0\arcmin$), selected as candidates
    by our PM XDGMM classifier. RR~Lyrae stars for each GC are
    overplotted, with RRab stars represented as blue triangles and RRc
    stars as green squares. Dubious candidates are highlighted with
    red circles. A complete version of this figure, showing all 26 VVV
    GCs with assigned RR Lyrae members, is provided in Fig.~\ref{fig_cmd1}.}
  \label{fig_cmd1short}
\end{figure*}

\subsection{Membership assignment}
\label{sec_membership}
We still needed to identify which of the VIVACE RR~Lyrae stars
detected in the vicinity of the VVV GCs are indeed members of the
clusters. To do this, we modified to some extent the procedure we
followed in \citetalias{alo21}.  We began by selecting all the VIRAC2
sources located inside the tidal radius of the different clusters in
our sample.  However, unlike in \citetalias{alo21}, we now rely on the
more recent coordinates and tidal radii $R_t$ for the Galactic GCs
provided in the 4th version of the Galactic Globular Clusters
Database\footnote{\url{https://people.smp.uq.edu.au/HolgerBaumgardt/globular/}}
\citep[GGCD;][]{bau18} as given in Table~\ref{tab_gcinput}.  Taking
the proper motions (PM) and corresponding errors of the sources for a
given cluster reported in VIRAC2, which include the VIVACE variable
stars, we used extreme deconvolution Gaussian mixture modeling
\citep[XDGMM;][]{hol17} to assign a probability of cluster membership
to every selected VIRAC2 source. Although in \citetalias{alo21} we
used a k-nearest neighbor (kNN) algorithm, in the present paper we
opted for the XDGMM algorithm, as the distribution of stars from the
GC and the field are well represented by such mixture models in the PM
space \citep{vas21}. This algorithm also allows us to incorporate PM
errors, which VIRAC2 provides.  We modeled the stellar PM distribution
with two Gaussian functions: the narrower one corresponds to the stars
in the GC and the wider one to the stars in the field.  The VVV GCs
show a broad range of masses and brightnesses (see
Table~\ref{tab_gcinput}), and the field density at the sky positions
of the different GCs varies significantly \citep{alo18}. However,
defining the training sample for the GCs to contain only stars close
to their centers ($r \le 0.75\arcmin-2\arcmin$) and with
well-determined VIRAC2 PMs
($\sigma_{\mu_{\alpha^\ast}}<1$~mas~yr$^{-1}$,
$\sigma_{{\mu}_{\delta}}<1$~mas~yr$^{-1}$), we obtained good
definitions for both field and cluster PM distributions in the VVV
GCs.

Unfortunately, for most of the VVV GCs the PM distribution of the
cluster stars shows some overlapping with that of the field stars,
which does not allow for a completely clean selection (see
Sect.~\ref{sec_pm}). Therefore, we need to use additional methods to
refine our membership assignment. Proximity to the cluster center also
increases the probability for a star to be a cluster member
\citep{alo11}.  However, keeping only RR~Lyrae stars close to the
cluster center would undermine our goal, which is to characterize all
the RR~Lyrae variables belonging to the VVV clusters.  As another
means to check on their GC membership, we can also use the RR~Lyrae
tight near-infrared PLZ relations, which once applied would result in
significant groupings in their distance moduli (see
Sect.~\ref{sec_distance2}).

To determine the GC membership of the sampled RR~Lyrae stars, we
define a method that makes use of all this information.  First, we
select only those that belong to a given VVV GC according to the PM
XDGMM algorithm (probability $> 0.5$), which are in the inner regions
of the cluster, no more than two half-light radii from its center
($r<2\,R_h$), and whose distance moduli, according to their PLZ
relations, agree within a 3-sigma clipping mean. Once this initial
approximation to the GC distance is established, we relax the
requirement about the projected distance to the GC center to include
all RR~Lyrae stars within the GC ($r<R_t$) fulfilling the other two
requirements. We report these RR~Lyrae stars as bona fide cluster
members in Sect.~\ref{sec_var}.

We also report as dubious cluster candidates those RR~Lyrae stars
located within the innermost regions of the GCs ($r<2\,R_h$) that
fulfill at least one of the other two requirements: membership based
on PM or based on the distance moduli derived from the PLZ
relations. Although in principle those pulsators should not be
considered as cluster members, we decided to consider them as
potential candidates as the photometry or astrometry of the stars
in the core of the GCs is sometimes not well defined in VIRAC2
due to the very high crowding of these regions \citep{maj12a,maj12b}.

Finally, for GCs with only one RR~Lyrae star located in the inner
regions ($r<2\,R_h$) and with appropriate PM, we consider these classical
pulsators as possible candidate members, as we are unable to apply
the membership procedure based on the PLZ relations.

\subsection{Distance determination}
\label{sec_distance}
As mentioned in the previous sections, we can find the distance to the
RR~Lyrae stars using their near-infrared PLZ relations. In
\citetalias{alo15}, we defined a set of PLZ relations calibrated in
the VVV filter system, which we slightly recalibrated in
\citetalias{alo21}. To use these relations, in addition to the
magnitudes and periods of the RR~Lyrae stars provided by VIVACE, we
also need their metallicities. We obtained them applying eq.~1 in
\citet{nav17}, which required the iron content $[{\rm Fe/H}]$ and the
${\alpha}$-enhancement [$\alpha$/Fe] of the RR~Lyrae stars. We assumed
a common metallicity for all the stars in a given GC. For the iron
content, we used the $[{\rm Fe/H}]$ values in Table~\ref{tab_gcinput}
derived mainly from \citet{sch24} based on APOGEE DR17, complemented
by other recent spectroscopic studies (see
Table~\ref{tab_gcinput}). As for the $\alpha$-enhancement, we assumed
an $[{\rm \alpha/H}]=0.3$ for all GCs.  The periods of the RRc stars
also needed to be fundamentalized using the relation
$\log~P_{\rm ab}=\log~P_{\rm c}+0.127$ \citep{del06,nav17}.

To properly calibrate our PLZ relations, we apply them to the RR~Lyrae
stars associated with NGC~6656 (M~22), and compare the results with
the distance provided by {\em Gaia} measurements in the GGCD. This
approach follows the method we used to recalibrate the PLZ relations
in \citetalias{alo21}. To align the two sets of distances for M~22, we
found we need to apply a slight offset $\Delta M=-0.026$ to the PLZ
relations from \citetalias{alo21}, transforming them into
   \begin{equation}
     M_{K_s} = -0.506 - 2.347 \log(P) + 0.1747 \log(Z),
   \end{equation}
   \begin{equation}
     M_H  = -0.423 - 2.302 \log(P) + 0.1781 \log(Z),
   \end{equation}
   \begin{equation}
     M_J  = -0.105 - 1.830 \log(P) + 0.1886 \log(Z),
   \end{equation}
   \begin{equation}
     M_Y  = +0.140 - 1.467 \log(P) + 0.1966 \log(Z),
   \end{equation}
   \begin{equation}
     M_Z  = +0.288 - 1.247 \log(P) + 0.2014 \log(Z).
   \end{equation}

 We chose M~22 as the reference because, as we detail in
 Sect.~\ref{sec_distance2}, this is the GC in our sample whose
 distance was obtained using the most methods from {\em Gaia} and {\em
   Hubble} Space Telescope (HST) measurements \citep{bau21} and
 also contains the highest number of references from the literature in
 the distance provided by the GGCD.

\section{CMDs and PM in the VVV GCs }
\label{sec_results}
The 41 GCs in the 4th version of the GGCD that fall in the footprint
of the VVV survey exhibit a diverse range of physical and observational
parameters (see Table~\ref{tab_gcinput}). In this section we will
present the CMDs and the PM vector-point diagrams (VPDs) of the stars
in the VVV GCs, focusing on the positions of the cluster RR~Lyrae
stars in them.

\begin{figure*}
  \centering
  \includegraphics[scale=1.0]{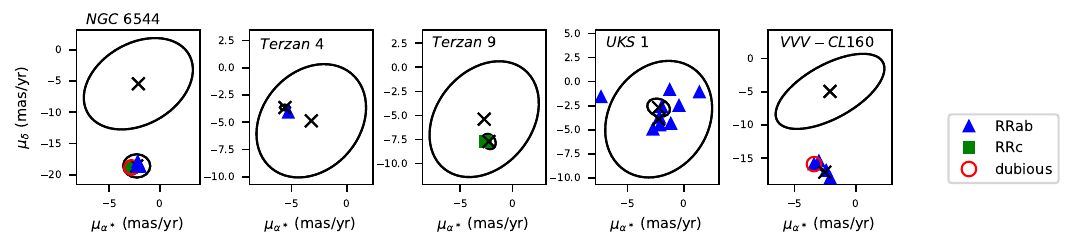}
  \caption{VPDs showing the PM of the RR~Lyrae cluster members of
    selected VVV GCs. RRab stars are represented as blue triangles,
    while RRc stars are shown as green squares. Red circles indicate
    dubious candidates. Ellipses representing the $2\sigma$
    distributions of field and cluster stars, as determined by our
    XDGMM algorithm, are overplotted, with crosses marking the centers
    of both distributions. A complete version of this figure, showing
    all 26 VVV GCs with assigned RR Lyrae members, is provided in
    Fig.~\ref{fig_pm}.}
    \label{fig_pmshort}
  \end{figure*}

\subsection{CMDs of the VVV GCs}
\label{sec_cmds}
Since it is constructed using PSF photometry, the VIRAC2 catalog
allows us to recover deeper and more complete near-infrared CMDs for
the VVV GCs \citep{alo18}. We present these CMDs in
Figs.~\ref{fig_cmd1short}~and~\ref{fig_cmd1} for the VVV GCs with
RR~Lyrae stars, and in Fig.~\ref{fig_cmd2} for those with no detected
RR~Lyrae stars in them. Due to the low Galactic latitudes where they
all lie, field star contamination is significant. To minimize this
effect in our CMDs, we plot only stars in the inner regions of the GCs
($r\le1\arcmin$) that are considered members according to the PM
analysis from our XDGMM algorithm (membership probability $>
0.5$). The short exposure times in the VVV survey makes our near
infrared VIRAC2 photometry deep enough to only reach the upper main
sequence above the magnitude confusion limit for two relatively nearby
clusters, NGC~6544 and NGC~6656 (M~22). However, we do reach the more
evolved regions of the red giant branch (RGB) and horizontal branch
(HB) for all of them. Therefore, we are covering with our VIRAC2
photometry the region where the RR~Lyrae stars are located in these
GCs, and VIVACE should have been able to identify them. Even in the
highly reddened, distant cluster UKS~1, VIVACE detects a significant
number of RR~Lyrae stars just over the detection limit. On the other
hand, for the aforementioned nearby NGC~6544 and M~22, VIVACE finds
RR~Lyrae stars just below the saturation limit (see
Figs.~\ref{fig_cmd1short}~and~\ref{fig_cmd1}).  There are 15 GCs in
our sample for which our analysis does not identify any member
RR~Lyrae stars (see Fig.~\ref{fig_cmd2}). Most of them are metal-rich
GCs that show very red HBs in their CMDs, which do not reach the bluer
colors of the RR~Lyrae stars. This is the case of Liller~1, NGC~6440,
NGC~6528, NGC~6553, NGC~6624, NGC~6637, Pal~6, Terzan~2, Terzan~6,
Terzan~12 and Ton~2, all of them with $[{\rm Fe/H}]>-0.92$ as shown in
Table~\ref{tab_gcinput}. The other GCs which do not present any
RR~Lyrae stars show very poorly populated CMDs in their inner regions,
where we expect to find most of these pulsating stars. This is the
case of Djorg~1, Gran~1, Gran~5, and VVV-CL001.  On the other hand, in
the other 26 GCs in our sample we were able to identify RR~Lyrae stars
(see Fig.~\ref{fig_cmd1}). These metal-poor GCs ($-1.70<[{\rm
    Fe/H}]<-0.78$) present CMDs that, in general, show well-populated
RGBs and blue HBs. But even in some cases where the CMDs are more
scarce in members, we could identify at least one RR~Lyrae candidate
(e.g., BH~261, Terzan~4, or Terzan~9).

\subsection{PM of the VVV GCs}
\label{sec_pm}
In our VIRAC2 dataset, the VPDs of the PM of the VVV GCs generally
show highly clumped distributions for cluster stars, superimposed on a
wider distribution of field stars, as observed in
Figs.~\ref{fig_pmshort}~and~\ref{fig_pm}.  Such PM distributions help
to detect field stars with PM incompatible with being cluster members;
however, for stars with PM compatible with being members, they do not
allow for a clear discrimination between cluster and field stars, even
at projected distances relatively close to the cluster centers as
field stars rapidly dominate over cluster members at these low
latitudes. Accurate PM separation is even more complicated for the VVV
GCs where the centers of the cluster and field distributions are very
close, as in NGC~6380, NGC~6441, NGC~6528, or NGC~6638. On the other
hand, a few VVV GCs has clearly separated PM distributions of cluster
and field stars (e.g., 2MASS-GC02, NGC~6544, NGC~6656, VVV-CL160),
which allows for a more reliable identification of the GCs' stellar
populations, including their RR Lyrae variables, extending to their
outer regions.

As a by-product of our cluster membership assignment, we obtained the PM
of the VVV GCs. In Table~\ref{tab_gcpm}, we present the PM we derived
by applying the XDGMM algorithm on the VIRAC2 data. We observe
good agreement with the GGCD data obtained from {\em Gaia}, except for UKS~1
and 2MASS-GC~02. These two GCs show some of the highest extinctions
among the VVV GCs, as we can infer from their very reddened CMDs (see
Fig.~\ref{fig_cmd1}) and will detail further in
Sect.~\ref{sec_distance2}. Therefore, observing these GCs at optical
wavelengths with {\em Gaia} is highly  challenging, and misidentification of
cluster members may have led to incorrect PM assignment. On the other
hand, the VIRAC2 near-infrared data from these GCs allow for a clearer
identification of their members. We previously observed this same
effect for 2MASS-GC~02 in \citetalias{alo21} and for UKS~1 in \citet{fer20}.

\section{RR~Lyrae stars in the VVV GCs}
\label{sec_var}
Even though all 41 VVV GCs have RR~Lyrae stars identified by VIVACE
inside their tidal radii, only in 26 of them we identified any bona
fide RR~Lyrae members after performing the analysis detailed in
Sect.~\ref{sec_analysis}. We summarize the number of detections of the
different types of RR~Lyrae stars in these 26 VVV GCs in
Table~\ref{tab_gcvar1} and compare them to the most updated version of
the Catalogue of Variable Stars in Globular Clusters
\citep[CVSGC;][]{cle01} or, when available, more recent references
from the literature. Table~\ref{tab_gcvar2} reports the main
observational parameters of the VVV GC RR~Lyrae stars as reported by
VIVACE, along with a flag reporting on their membership assignment
based on the analysis in Sect.~\ref{sec_analysis}.  We present in
Figs.~\ref{fig_cmd1short}~and~\ref{fig_cmd1} the CMDs of the VVV GCs
with the positions of the detected RR~Lyrae stars
superimposed. Additionally, we also present the VPDs of the PM of the
cluster RR~Lyrae stars, along with the distributions of the field and
cluster populations, in Figs.~\ref{fig_pmshort}~and~\ref{fig_pm}.  In
the following subsections we discuss in more detail the
characteristics of the detected RR~Lyrae stars in the individual VVV
GCs. In particular, we emphasize the comparison of our membership
assignments with those from the recent studies by \citet{cru24} and
\citet{pru24}, which use the third data release (DR3) from the {\em
  Gaia} mission to characterize and assign membership to RR~Lyrae
stars in the Galactic globular clusters.

\subsection{2MASS-GC02}
\label{sec_2ms}
We previously explored the variable stars in 2MASS-GC02 in
\citetalias{alo15} and \citetalias{alo21}. VIVACE lists all the
RR~Lyrae stars that we found in those papers.  Our analysis identifies
16 RRab and two RRc as bona fide cluster members. All the RRab
considered cluster members in \citetalias{alo15} and
\citetalias{alo21} are identified, plus two new RRab (1020301 and
1020311)\footnote{The identification numbers for the RR~Lyrae stars
referenced hereafter correspond to their VIVACE IDs.} and also, for
the first time, two RRc (1020312 and 1020316) in this GC. We include
them in Tables~\ref{tab_gcvar1}~and~\ref{tab_gcvar2} as bona fide
members. There is one more VIVACE RRab, 1020332, which lies near the
cluster center ($R<1.58\arcmin=1.1\,R_h$), with a high probability of
being a cluster member according to our PM analysis. However, the
distance and color inferred from the PLZ relations do not agree with
those of the other RR~Lyrae stars, so we include it as a dubious
candidate in Tables~\ref{tab_gcvar1}~and~\ref{tab_gcvar2}. There are
13 RRab in the CVSGC for this GC, taken from our
\citetalias{alo15}. We considered 12 of them as cluster members in
both \citetalias{alo15} and \citetalias{alo21}, but we pointed out in
both papers that V31, the RRab farthest from the cluster center in
projection, should be considered a field star. Our current analysis
further confirms this fact, and we flagged this star in
Table~\ref{tab_gcvar2} accordingly. Neither \citet{cru24} nor
\citet{pru24} include 2MASS-GC02 in their analysis, due to its high
reddening.

\subsection{BH~261 (AL~3)}
\label{sec_bh261}
BH~261 does not appear in the CVSGC, but in the recent study of this
GC by \citet{kun24}, they found one RRc star that they considered a
member. \citet{cru24} also identify this pulsator in their Gaia sample
as a cluster member. VIVACE lists this variable star, alongside other
pulsators within the tidal radius of this GC. However, after applying
our XDGMM algorithm to the PM, only 778882, the RRc found by
\citet{kun24} and \citet{cru24}, is assigned to BH~261. This is also
the closest pulsator to the cluster center ($r=0.37\arcmin$), and the
only one within its half-light radius, which reinforces its
probability of belonging to the cluster.

\subsection{Djorg~2}
\label{sec_djorg2}
The CVSGC lists five RRab and two RRc for Djorg~2,
taken from the OGLE variable star database \citep{sos14} based on their
relative distance to the cluster center ($r<2\arcmin$). VIVACE only detects
three of these RRab: V1=784125, V2=784166, and V5=786666.  These are the
only VIVACE pulsators for this GC that our analysis considers as
cluster members. \citet{cru24} also identify these three pulsating stars as cluster members, but \citet{pru24} only recognize V2 as a member in their study.

\subsection{FSR~1716 (VVV-GC05)}
\label{sec_fsr1716}
FSR~1716 does not appear yet in the CVSGC, but its variable population
has been recently explored by our collaboration. \citet{min17a} found
12 RRab that they tentatively assigned to the cluster based on their
projected proximity. \citet{con18b}, using PM analysis, refined the
assignment, keeping six RRab and three RRc as cluster members. VIVACE finds
all of these variable sources, but our PM analysis only considers five
RRab and three RRc as members, all included in the \citet{con18b}
selection. These eight RR~Lyrae stars are also the closest to the GC
center, with $r<2\,R_h$, and their PLZ relations suggest a common
distance (see Fig.~\ref{fig_dis1}). We include them in
Tables~\ref{tab_gcvar1}~and~\ref{tab_gcvar2} as bona fide members.
However, although they are all included in the study by \citet{cru24},
it is worth noting that they are classified as non-members in their
analysis. Additionally, we point out that the three variable stars from
\citet{min17a} that both \citet{con18b} and we disregard as cluster
members are not even considered RR~Lyrae stars but eclipsing binaries
by VIVACE. There is also another RRab in \citet{min17a},
1231449=d025-0114911, which our XDGMM algorithm disregards as a
cluster member, while \citet{con18b} do not, based on its
PM. \citet{cru24} do not list it in their sample.  This RRab is much
further from the cluster core ($r<11.6\arcmin=6.8\,R_h$) and its
period is much shorter than that of the other RRab members, but its
PLZ relations suggest a common distance modulus with the other
RR~Lyrae stars in the cluster. However, based on the selection
criteria we listed in Sect.~\ref{sec_analysis}, we do not consider it
a member.

\subsection{FSR~1735 (2MASS-GC03)}
\label{sec_fsr1735}
FSR~1735 does not appear yet in the CVSGC, but in our VVV
collaboration, \citet{car16} explored this GC and found three candidate
RRab in its neighborhood. VIVACE identifies all these RR~Lyrae
stars\footnote{There is an erratum in the reported declination of V2
in \citet{car16}, which should be $-47\fdg00717799$}, but our analysis
considers as cluster members only V1=1129841 and V2=1129823, the two
closest pulsators to the GC center ($r<1.2\,R_h=1.2\arcmin$).  We
include them in Tables~\ref{tab_gcvar1}~and~\ref{tab_gcvar2} as bona
fide members. Our analysis discards V3=1129230, which in addition to
being located far from the center ($r\sim8\,R_h=8.2\arcmin$),
has a low probability of being a cluster member according to the PM
analysis. In their study, \citet{cru24} classify only V1 as a cluster member.

\subsection{HP~1 (FSR~1781)}
\label{sec_hp1}
Our analysis identifies six RRab and three RRc as bona fide cluster members
of HP~1, which are listed in
Tables~\ref{tab_gcvar1}~and~\ref{tab_gcvar2}. Additionally, we
classify one RRc, 290373, as a dubious member. Although it lies near the
cluster center ($r\sim1.2\arcmin=0.8\,R_h$) and its distance modulus,
based on the PLZ relations, is consistent with that of the other
RR~Lyrae members (see Fig.~\ref{fig_dis1}), the XDGMM analysis of the
PM assigns it a low membership probability. We have flagged this star
accordingly in Table~\ref{tab_gcvar2}.  The CVSGC does not list any
RR~Lyrae stars for HP~1. However, \citet{ker19} provided a
comprehensive study of this GC, including its variable
population. Taking the RR~Lyrae stars that are closer than
$150\arcsec$ from the cluster center in the OGLE catalog
\citep{sos14}, they found six RRab and five RRc compatible with being
cluster members. VIVACE detects all but two RRc stars from their
selection. Our analysis, however, does not classify two RRab,
OGLE-BLG-RRLYR-19433=290346 and OGLE-BLG-RRLYR-19503=290374, as
members, and we have flagged them as non-members in
Table~\ref{tab_gcvar2}. \citet{vas21b}, in her M.Sc. thesis, also
studied the variable population of HP1 using VVV observations. She
reported 18 variable stars as potential cluster members, including eight
RRab and four RRc stars. VIVACE misses one RRab (HP1-9) and one RRc (HP1-10)
from her list. And while our analysis considers one of her listed
RRab, HP1-17=290229, as a field star, we confirm two RRab (HP1-14=290365
and HP1-5=290369) as bona fide members previously unreported by
\citet{ker19}. Lastly, \citet{cru24} also listed six RRab and five RRc as
HP~1 members. VIVACE misses one RRc, the same as in the two previously
cited studies, but our analysis confirms the cluster membership of
another RRc (290366), which had not been reported by either
\citet{ker19} or \citet{vas21b}.

\subsection{NGC~6380 (Ton~1)}
\label{sec_ngc6380}
Our analysis assigns seven RRab and three RRc as bona fide members of
NGC~6380, and one additional RRab as a dubious member.  We list them in
Tables~\ref{tab_gcvar1}~and~\ref{tab_gcvar2}. There are no RR~Lyrae
stars associated with NGC~6380 according to the CVSGC,
but using OGLE data, \citet{sos19} consider as members the nine RR~Lyrae
stars inside their considered cluster radius ($r<3.6\arcmin$). The
region we explore, extending out to the tidal radius
($R_t=14.4\arcmin$, see Table~\ref{tab_gcinput}) is significantly
larger. We emphasize that we classify as bona fide cluster members all
the inner RR~Lyrae stars from the OGLE catalog cited by \citet{sos19}
for this GC, except for OGLE-BLG-RRLYR-51429, which is the closest to
the cluster center and that VIVACE fails to identify. \citet{cru24}
also classifies as cluster members the three RRab and three RRc in
common. Adding to those, there are three RRab (85269, 85270, and 90714) in
the outskirts of NGC~6380 ($4.1\arcmin\le r\le 12.4\arcmin$) that we
classify as cluster members. The RRab 85269 is also classified by
\citet{cru24} as a cluster member, but the other two are newly
discovered cluster candidates. The farthest of these from the cluster
center, 90714, has a borderline probability (46\%) of belonging to the
GC according to the PM analysis, so we flag it as a dubious candidate.

\subsection{NGC~6401}
\label{sec_ngc6401}
NGC~6401 is rich in RR~Lyrae stars. Our analysis considers 21 RRab and
three RRc in VIVACE as bona fide members of NGC~6401, plus two more RRab and
one RRc as dubious members. They are all included in
Tables~\ref{tab_gcvar1}~and~\ref{tab_gcvar2}. The CVSGC for NGC~6401
reports 23 RRab and 11 RRc, taken from the recent
variability study by \citet{tsa17}. VIVACE fails to detect four RRab and
seven RRc, most of which are located in the very core of the
cluster. Among the 19 RRab and four RRc that VIVACE detects in common
with the CVSGC, all of them are considered cluster members by our
analysis, except for one RRab, V17=564279, which has been flagged in
Table~\ref{tab_gcvar2} as a field star. However, there are two RRab in
this group (V32=563533 and V34=564266) whose position in the very core
of the cluster and PM probabilities suggest they are cluster members,
but whose distance moduli are $\sim0.2$ mag smaller than the others.
However, their mean $K_s$ magnitude reported in VIVACE may have easily
been affected by the high crowding of the core.  Likewise, there is
also another RRc, V23=564278, whose position near the cluster core
($r=2.2\arcmin\sim2.1\,R_h$) and distance modulus according to the PLZ
relations put it inside the cluster (see Fig.~\ref{fig_dis1}), but
whose PM do not agree with those of NGC~6401, raising a question about
its membership. We report these three RRab as dubious cluster member
candidates in Tables~\ref{tab_gcvar1}~and~\ref{tab_gcvar2}.
Additionally, we include five more RRab (563525, 564296, 558881, 558882,
and 564302) as bona fide cluster members which do not appear in the
CVSGC. The reason for their omission may have been their location
outside the radius $R=2.4\arcmin$ where \citet{tsa17} focused their
search for cluster members. These five pulsating stars, however, were
identified and classified as cluster members in \citet{ara20}, who
used VVV data in his M.Sc. thesis to study NGC 6401, and later in
\citet{cru24}. We note the case of the RRab 564302. The period
reported by VIVACE appears to be aliased, as it is close to twice the
value reported by OGLE, {\em Gaia}, and \citet{ara20}. We consider it
a bona fide cluster member, as its PM analysis shows a high
probability of membership and the PLZ relations, when including the
recalculated period, places this star at a distance consistent with
the other RR~Lyrae members. Finally, we note that \citet{ara20} agrees
with our membership assignments for all common RR~Lyrae
stars. However, for \citet{cru24} and \citet{pru24} there are a few
stars for which our membership assignments differ. Differently
from our analysis, \citet{cru24} classify V17 as a cluster member and
V16, V22, V29, and V32 as field stars, while \citet{pru24} consider
V2, V6, V23, V26, V32, and V34 as field stars.

\subsection{NGC~6441}
\label{sec_ngc6441}
Along with NGC~6388, NGC~6441 forms the Oosterhoff III group of
high-metallicity GCs that contain a significant number of long-period
RR~Lyrae stars \citep{pri00}. Our analysis identifies 20 RRab and five
RRc as bona fide members of NGC~6441 among the VIVACE variables, plus
two additional RRab and one RRc as dubious members. These stars are all
included in Tables~\ref{tab_gcvar1}~and~\ref{tab_gcvar2}.

The CVSGC for NGC~6441 lists 44 RRab (plus three more possible RRab), 19
RRc (plus seven more possible RRc), one RRd, and two RR~Lyrae with no subtype
classification.  This high number of RR~Lyrae stars was discovered
mainly in the innermost GC by \citet{pri03}, who took full advantage
of the superb resolution of the HST. Our ground data misses most
of the stars located in the innermost GC regions, detecting only two
pulsators of the 55 listed at $r<1.0\arcmin$. However, VIVACE detects
most of the cluster members at $r>1.0\arcmin$, missing only two RRab
(V41 and V97) and three RRc (V71, V74, and V102). We note that the
coordinates of two RRab members (V40 and V59) differ by more than
$1\arcsec$ between CVSGC and VIVACE.  Moreover, there are five additional
stars classified as RR~Lyrae members in the CVSGC but that VIVACE
classifies as eclipsing binary stars (V52, V78, V93, V96, and
V150). After examining their light curves, we considered these as
misclassifications from VIVACE and recalculate the period following
\citetalias{alo21}. We treat them as bona fide RR~Lyrae members, as
their PM and distances derived from the PLZ relations are consistent
with those of other RR~Lyrae members. Similarly, V69 is classified as
an RRab by VIVACE and as an RRc in the CVSGC. Although its period
seems to match an RRab, its low amplitude suggests it is an RRc. We
retained this classification, as it places the star at the GC distance
based on the PLZ relations. We had a similar discussion about this
star in \citetalias{alo21}. All these variable stars, for which we
modified their variability classification, are flagged in
Table~\ref{tab_gcvar2}.

The CVSGC also lists three RRab (V36, V54, and V67) and two RRc (V68 and
V73) as possible field members in NGC~6441. VIVACE detects them as
well, and our analysis confirms that they are not cluster members. We
have flagged these stars in Table~\ref{tab_gcvar2}.  There are three more
RR~Lyrae stars from the CVSGC whose distance moduli from the PLZ
relations set them as field stars, although they could be members
according to PM and distance to the center: V45, V49, and V70. As we
noted earlier, PM separation between cluster and field stars for
NGC~6441 is complicated, so we prefer to consider them as field
variable stars, as we did in \citetalias{alo21}. On the other hand, one
RRab (V53) and one RRc (V77) are located in the inner regions of the
cluster, and their distance moduli based on the PLZ relations place
them as cluster members, but their XDGMM probabilities are low. We
include them as dubious members in
Tables~\ref{tab_gcvar1}~and~\ref{tab_gcvar2}.

We previously studied NGC~6441 in \citetalias{alo21} and reaffirm our
membership assignments for all the stars in common, including an
additional RRab, C36=105687, which was first reported as a member in
\citetalias{alo21}. We also agree with the membership assignment of
\citet{oli22} for all the RR~Lyrae stars in common, except for V45,
V66, V70, and V72. They also report a newly classified RRc member,
OGLE-BLG-RRLYR-30179, which is not detected by VIVACE.  \citet{cru24}
list one additional new RRab member, Gaia-4040152937984200576, but
although VIVACE detects it (105236), our analysis considers it a field
star.  For the stars in common, all our bona fide members are also
considered cluster members by \citet{cru24}, except V40, V42, and
V66. On the other hand, all of our field RR~Lyrae stars are considered
cluster members by \citet{cru24}, except V70. For our two dubious
members in common, \citet{cru24} classify V77 as a member, but
considers V53 a field star. We also compared our membership
assignments with those of \citet{pru24} and found disagreements for a
significant number of RR~Lyrae stars: V37, V39, V43, V51, V53, V62,
V66, V70, V72, V77, and V93.

Finally, we highlight that we consider three RRab (105175, 105856, and
198097) and one RRc (105528) as newly classified members of NGC~6441.
While all of them are considered members according to their PM by our
XDGMM classifier and the PLZ relations set them at the right distance,
we notice that they are quite far from the cluster core
($16.6\arcmin\le r\le 25.9\arcmin$), which is probably the reason for
having been missed before. Since one of these RRab, 198097, has a
borderline probability (44\%) of belonging to the GC according to the
PM analysis, we flagged it as a dubious candidate.

\subsection{NGC~6453}
\label{sec_ngc6453}
The CVSGC lists three RRab and five RRc for NGC~6453. As with
other GCs, VIVACE misses most of the variables in the inner regions
($r<1\arcmin$), but detects the majority of the pulsators that are
beyond. Unfortunately, this leaves us with only one RRab, V8=207475,
which our analysis classifies as a bona fide cluster member. There are
other two pulsators, the RRab V4=207471 and the RRc V3=207472, near
the center of the cluster ($r<1.5\arcmin=1.6\,R_h$) and located at the
same distance as V8 according to the PLZ relations (see
Fig.~\ref{fig_dis1}), but having low probabilities of belonging to the
cluster according to their PM. Thus, we classify them as dubious
cluster members in Tables~\ref{tab_gcvar1}~and~\ref{tab_gcvar2}. Both
\citet{pru24} and \citet{cru24} classify V4 and V8 as cluster members,
but consider V3 a field star. VIVACE also detects two additional CVSGC
RRc, V5=207460 and V6=207459, but our analysis considers them field
stars. \citet{pru24} agree with this classification, while
\citet{cru24} classify them as cluster members.

\subsection{NGC~6522}
\label{sec_ngc6522}
Our analysis classifies four VIVACE RRab and three VIVACE RRc as bona fide
members of NGC~6522. We list them in
Tables~\ref{tab_gcvar1}~and~\ref{tab_gcvar2}.

According to the CVSGC for this cluster, it harbors five 
RRab and six RRc. VIVACE fails to identify two RRab (V10 and V11) and two
RRc (V12 and V14), likely due to their very central location
($r<0.3\arcmin$ for V10, V11 and V14). The rest of the CVSGC RR~Lyrae
stars for NGC~6522 appear in VIVACE and we classify all but one RRc,
V9=735496, as cluster members (see
Table~\ref{tab_gcvar2}). Furthermore, we identify one RRab member,
V4=736543, with an offset of $2\arcsec$ from the position provided in
the CVSGC, as also reported by \citet{are23} in their appendix. There
are six additional RR~Lyrae stars reported in the CVSGC notes for
NGC~6522 as found in the OGLE catalog in the vicinity of the cluster
with magnitudes appropriate for cluster membership. VIVACE identifies
four of them, but according to our PM analysis they have low
probabilities of being cluster members.

Recently, \citet{are23} revisited the variable population of
NGC~6522. For the stars in common, our membership assignments agree
with theirs, including the OGLE stars listed in the CVSGC notes. They
also report one additional RRc, V24, as a new member, but unfortunately,
VIVACE fails to detect it. \citet{cru24} also report one new RRab
member, Gaia-4050201237330707712. VIVACE detects this star (737755), but our
analysis classifies it as a field star, in agreement with
\citet{are23}. For the remaining stars in common with \citet{cru24},
our membership assignments agree with theirs, except for V4 and V9. In
the case of \citet{pru24}, our membership assignment only differs for
V2.  Finally, we highlight the discovery of a new RRab member, 735495,
in our data.

\subsection{NGC~6540}
\label{sec_ngc6540}
There are two RRab and one RRc listed as members in the 2016 update of
the CVSGC for NGC~6540, first discovered and assigned to the cluster
by the OGLE collaboration \citep{sos11}. VIVACE detects one of the RRab,
V3=796385, and the RRc, V1=796378, and our analysis considers them as
cluster members. However, we note that both \citet{cru24} and
\citet{pru24} consider these two pulsators as field stars. VIVACE finds
another RRab, 796362, close to the cluster core
($r\sim1.5\arcmin=R_h$), which we highlight as a new bona
fide cluster member.  We include it along with the other two CVSGC RR~Lyrae
stars in Tables~\ref{tab_gcvar1}~and~\ref{tab_gcvar2}.

\subsection{NGC~6544}
\label{sec_ngc6544}
Our analysis identifies three RRab in NGC~6544 as bona fide cluster
members and one RRc as a dubious member, all of which are listed in
Tables~\ref{tab_gcvar1}~and~\ref{tab_gcvar2}. The relative proximity
of this GC, inferred from its CMD (see Fig.~\ref{fig_cmd1}), along
with its clear separation from the field in its VPD (see
Figs.~\ref{fig_pmshort}~and~\ref{fig_pm}), makes it particularly
straightforward to identify the members of NGC~6544 among the VIVACE
RR~Lyrae stars. Only one RRab is listed for this GC in the CVSGC, which
we also identify and classify as a cluster member, in agreement with
\citet{cru24} and \citet{pru24}. \citet{cru24} also identify one
additional RRab member in NGC~6544, Gaia-4065746094009651328. VIVACE
detects this star (908712), and we also classify it as a bona fide
member. Finally, we highlight the new identification of one additional
RRab, 881126, and one RRc, 919762, as members of NGC~6544. The distance
modulus to this RRc, based on the PLZ relations, is $\sim0.25$ mag
smaller than that of the other cluster RR~Lyrae stars (see
Fig.~\ref{fig_dis1}), leading us to consider it a dubious
candidate. \citet{cru24} also identify this star but classify it as a
field star.

\subsection{NGC~6558}
\label{sec_ngc6558}
The CVSGC lists seven RRab and three RRc as members for NGC~6558. Since three of
these stars, V9, V13 and V14, are located beyond the tidal radius for
this GC (see Table~\ref{tab_gcinput}), we extended our search radius
for RR~Lyrae stars to $r=1.6\,R_t=6.1\arcmin$ to include them. VIVACE
detects all of these stars, except V1, which is the closest to the
cluster center ($r\sim0.4\arcmin$). Our analysis classifies all these
RR~Lyrae stars as cluster members, except for V13=730703 and
V14=731330, the two stars farthest from the GC center. We flag them in
Table~\ref{tab_gcvar2}, along with V5=730699, an RRab listed by the
CVSGC as a possible field star, which we also detect and classify as a
non-member variable star. All the other RR~Lyrae stars (six RRab and one
RRc) are considered bona fide members of NGC~6558, and listed in
Tables~\ref{tab_gcvar1}~and~\ref{tab_gcvar2}. Notably, this includes
V9=730641, which lies outside the tidal radius of the cluster
($r=5.9\arcmin=1.5\,R_t$).

When comparing our membership assignments with recent studies, we find
a general agreement for most stars in common, with a few exceptions:
V8 and V9 with \citet{are24}; V4 with \citet{pru24}; and V5, V13, and
V14 with \citet{cru24}. Finally, there is an additional RRab detected
by VIVACE, 730683, with no counterpart in the CVSGC.  Although it has
low probability of being a cluster member according to our PM
analysis, it lies relatively close to the GC center
($r\sim1.1\arcmin=1.2\,R_h$) and its distance modulus, based on the
PLZ relations, agrees with those of the other RR~Lyrae members (see
Fig.~\ref{fig_dis1}). We include it in
Tables~\ref{tab_gcvar1}~and~\ref{tab_gcvar2} as a dubious
member. However, we note that only \citet{are24} list it (O47), but
classify it as a field star.

\subsection{NGC~6569}
\label{sec_ngc6569}
Our analysis identifies seven RRab and three RRc in NGC~6569 as bona fide
cluster members. We report them in
Tables~\ref{tab_gcvar1}~and~\ref{tab_gcvar2}.

The CVSGC for NGC~6569 associates nine RRab, 12 RRc,
and three possible additional RR~Lyrae stars with this GC.  VIVACE detects
four of the RRab, three of the RRc, and one of the possible RR~Lyrae,
V5=731452, which it classifies as an RRab.  Our analysis considers all
of these as cluster members.  Most of the remaining missed pulsators
from the CVSGC are located very close to the central region of the
cluster. VIVACE also detects two RR~Lyrae stars, V6 and V19, which the CVSGC
classifies as field stars. Our analysis also categorizes these as
non-members, and we flag them accordingly in
Table~\ref{tab_gcvar2}. Notably, these are the RR~Lyrae stars farthest
from the cluster center.

In \citetalias{alo21}, we previously studied the variable population
of this cluster. VIVACE detects all but one RRab and one RRc reported in
that paper. It also detects two additional RRab, 731451 and 731496,
which our analysis classifies as cluster members, previously reported
as such in \citetalias{alo21} (variables C2 and C13, respectively),
but which are still unreported in the CVSGC. \citetalias{alo21} agrees
with our membership assignments for all common stars, while
\citet{cru24} disagrees for V5, V6, V19, and V20. \citet{pru24} also
agree for all stars in common, except for V5, V20 and V26.

\subsection{NGC~6626 (M~28)}
\label{sec_ngc6626}
We identify eight RRab and six RRc as bona fide members of NGC~6626,
along with two additional RRab and one RRc as dubious members. We include
them all in Tables~\ref{tab_gcvar1}~and~\ref{tab_gcvar2}.

According to the CVSGC, this GC hosts
ten RRab and eight RRc. VIVACE misses only four RRc, all of which were newly
identified by \citet{pri12}. Both the OGLE and {\em Gaia} variable star
catalogs also fail to identify three of these as variables. Our analysis
confirms the membership of the remaining RR~Lyrae stars in the CVSGC,
with the exception of three RRab (V9=969821, V18=969858, and V20=969859),
which have low probabilities of being cluster members according to
their PM. However, we classify V18 and V20 as dubious members, given
their proximity to the cluster center ($r<0.9\arcmin\sim0.9\,R_h$) and
their distance moduli, which, according to the PLZ relations, align
with those of the other RR~Lyrae members. Additionally, we note that two
of these RRab, V20=969859 and V25=969846, show an offset of more than
$1\arcsec$ compared to the positions reported in the CVSGC. VIVACE
detects three additional RRab stars (V15, V16, and V24), which are
classified as field stars by the CVSGC. Our analysis also categorizes
them as non-cluster stars, and we flag them accordingly in
Table~\ref{tab_gcvar2}. Notably, these field RRab are located farther
from the GC in projection than the confirmed RR~Lyrae members.

We previously analyzed the variable population of NGC~6626 in
\citetalias{alo21}, and identified one more RRab (969862) and three
additional RRc (969825, 969826, and 969852) as cluster members. VIVACE
also detects these stars, and our analysis confirms their membership
in the cluster. Although one RRc, 969826, has low probability of
belonging to the GC according to the PM analysis, its position near
the cluster center ($r\sim2\,R_h=2.1\arcmin$) and similar distance
modulus to the other cluster pulsators, according to the PLZ relations
(see Fig.~\ref{fig_dis1}), led us to consider it a probable cluster
member. We include it as such in
Tables~\ref{tab_gcvar1}~and~\ref{tab_gcvar2}. These four RR~Lyrae stars are
also identified as cluster variables in recent studies by
\citet{oli22} and \citet{cru24}. For the other pulsating stars in
common with our analysis, \citet{oli22} agree with our membership
assignment, except for one RRc, V9. In contrast, \citet{cru24} show more
disagreements, including V5, V9, V15, V16, V20, V24, and V25. We also
compare our membership assignments with those of \citet{pru24}, and
find differences only for three stars: V5, V22, and V25.

\subsection{NGC~6638}
\label{sec_ngc6638}
Our analysis identifies ten RRab and seven RRc as bona fide members of
NGC~6638. We list them in Tables~\ref{tab_gcvar1}~and~\ref{tab_gcvar2}.

The CVSGC reports ten RRab and 16 RRc, along with two additional RRc
classified as probable field stars for this cluster. VIVACE detects six
RRab and eight RRc from this list. We note that V32=1042563 is classified
as an eclipsing binary by VIVACE but as an RRab by the CVSGC. After
reviewing its light curve, position in the CMD, and distance modulus
derived from the PLZ relations, we could not definitively classify its
variable type. Its position, very close to the cluster center
($r=0.11\arcmin$), may introduce offsets in its photometry. Therefore,
we have flagged it as a dubious variable member in
Table~\ref{tab_gcvar2}, but excluded it from
Table~\ref{tab_gcvar1}. For the remaining RR~Lyrae stars in common,
our analysis considers all as cluster members, except for one RRc,
V38=1042556, which is classified as a field star. The two suspected
field RRc identified by the CVSGC, V40=1042543 and V43=1042547, which
are also detected by VIVACE, are classified as bona fide members in
our analysis. Furthermore, VIVACE classifies as RRab six variables of
unknown classification in the CVSGC. Among them, the three RRab closest to
the GC center, with $r<0.5\arcmin$ (V34=1042564, V35=1042562, and
V36=1042561), are classified as cluster members in our analysis, while
the other three RRab (V8=1042523, V51=1042524, and V52=1042527), which are
significantly farther away ($r>7.8\arcmin$), are classified as field
stars. They are flagged accordingly in Table~\ref{tab_gcvar2}.

The variable population of NGC~6638 has been recently revisited by
\citet{oli22}, who identified two new RRab members in the OGLE dataset,
OGLE BLG-RRLYR-62076 and OGLE BLG-RRLYR-62131, and by \citet{cru24},
who added one additional RR~Lyrae member,
Gaia-4076175962610552448. VIVACE detects one of the new RRab members
from \citet{oli22}, 1042546=OGLE BLG-RRLYR-62131, which our analysis
also classifies as a bona fide cluster member. In comparing all
RR~Lyrae stars in common, our membership assignments align with those
of \citet{oli22} in all cases, but disagree with \citet{cru24} for
V27, V30, and V38. Additionally, we find only two discrepancies with
the assignments of \citet{pru24} for V35 and V41.

Finally, we highlight our analysis includes one RRab, 1042542, as a
newly identified member of NGC~6638.

\subsection{NGC~6642}
\label{sec_ngc6642}
Our analysis identifies eight RRab and eight RRc as bona fide members of
NGC~6642, along with two additional RRab and two RRc as dubious
members. We list them in Tables~\ref{tab_gcvar1}~and~\ref{tab_gcvar2}.
In the literature we note several variable stars have been detected
beyond the tidal radius reported in Table~\ref{tab_gcinput}, which
prompted us to extend the studied region out to
$r\sim1.1\,R_t=7.6\arcmin$ to include these pulsators (see
Table~\ref{tab_gcvar2}).

According to the CVSGC, NGC~6642 contains seven RRab (plus three suspected), three
RRc (plus three suspected), and one probable RR~Lyrae of unclassified
subtype, belonging to this GC. VIVACE detects all of the RR~Lyrae
stars cataloged in the CVSGC. All suspected RRab and RRc in the CVSGC
are confirmed by VIVACE, and the RR~Lyrae star with a previously
unknown subtype is reclassified as an RRab. We observe an offset of
more than $1\arcsec$ in the coordinates of V6=1046390 and
V14=1046403. Additionally, we note that V4=1046405 is classified as an
RRab by VIVACE but as an RRc by the CVSGC, as well as by the OGLE and
{\em Gaia} catalogs. The period reported by VIVACE appears to be
aliased, as it is three times the value provided by the other
catalogs. After examining its light curve, position in the CMD, and
distance modulus derived from the PLZ relations, we consider it as a
misclassification by VIVACE and we classify this star as an RRc
member, recalculating its period as described in \citetalias{alo21},
(see Table~\ref{tab_gcvar2}). Our analysis categorizes all of the
RR~Lyrae star in the CVSGC for NGC~6642 as bona fide members, except
for V8, V11, V17, and V18, which are classified as field stars, and
V9, an RRab which is considered a dubious member due to its low
membership probability based on its PM.

Using OGLE detections, \citet{oli22} recently studied this cluster and
identified two new RRc members\footnote{There is one additional
high-probability RRab member in \citet{oli22} with no CVSGC
counterpart, OGLE-BLG-RRLYR-62397, but we believe it to be V6 based on
its coordinates and reported period.}, OGLE-BLG-RRLYR-36604 and
OGLE-BLG-RRLYR-62216. \citet{cru24} classified one additional new RRab
member, Gaia-4077768914409054464. VIVACE detects all three pulsating stars
(1046355, 1046368, and 1046386). Our analysis considers the two newly
detected RRc as bona fide members, while the new RRab is classified as
a dubious candidate, as its distance modulus, derived from the PLZ
relations, shows a discrepancy with those of the other RR~Lyrae
members. When comparing all RR~Lyrae stars in common, our membership
assignments generally agree with both studies. Discrepancies occur
only for V14 and OGLE-BLG-RRLYR-62390=1046394 with \citet{oli22}, and
for V8, V11, V17 and V18 with \citet{cru24}.  We also align with most
membership assignments by \citet{pru24}, except for V9 and V12.

Finally, we highlight the discovery of one bona fide new RRab member
(1046387), one bona fide new RRc member (1046394), and two dubious RRc
members (1046382 and 1046383) in NGC~6642.

\subsection{NGC~6656 (M~22)}
\label{sec_ngc6656}
As with NGC~6544, the relative proximity of NGC~6656, evidenced by its
bright sequences in the CMD (see Fig.~\ref{fig_cmd1}) and the clear
separation between cluster and field stars in its VPD (see
Fig.~\ref{fig_pm}), makes the identification of cluster members
particularly straightforward. Our analysis identifies ten RRab and 13
RRc as bona fide members of NGC~6656, which are listed in
Tables~\ref{tab_gcvar1}~and~\ref{tab_gcvar2}.

The CVSGC reports 11 RRab and 15 RRc as members of NGC~6656. All RRab
are detected by VIVACE, but five RRc are missed.  Additionally, there are
some discrepancies in the classification of variable types. V21 is
classified as an RRc by VIVACE and OGLE, but as an RRab in the
CVSGC. We maintain its classification as an RRc based on its short
period, small amplitude, and distance modulus from the PLZ relations
(see Fig.~\ref{fig_dis1}). V37 is classified as an RRab by VIVACE, but
upon analyzing its light curve, we consider it an eclipsing binary, as
suggested by the CVSGC and OGLE. Ku-1 is classified as an eclipsing
binary by VIVACE, but after reviewing its light curve and
recalculating its period following \citetalias{alo21}, we align it
with the RRc classification provided by both the CVSGC and OGLE. Upon
submitting the VIVACE-CVSGC cross-matched RR~Lyrae stars to our
membership analysis, we find that all of them are bona fide cluster
members.  It is also worth noting that NGC~6656 was studied in
\citetalias{alo21}, where we identified the same member RR~Lyrae
stars, plus an additional RRc, Ku-4, which VIVACE fails to detect. The
CVSGC for this GC also lists five RRab, one RRc and one RR~Lyrae star with an
unclassified subtype as field stars. VIVACE detects all of these stars
and classifies V29=1047961, the RR Lyrae star with an unknown subtype,
as an RRc, in agreement with OGLE and \citetalias{alo21}. Our analysis
also considers all the RRab as field stars, but classifies the two RRc,
V27 and V29, as bona fide cluster members, consistent with our
\citetalias{alo21} classification.

\citet{cru24} do not assign any new RR~Lyrae variables as members. Our
membership assignments differ from theirs only for two RR~Lyrae stars,
V3 and V22. On the other hand, \citet{pru24} agree with all our
membership assignments.

\subsection{Terzan~1 (HP~2)}
\label{sec_ter1}
Our analysis identifies seven RRab and two RRc as bona fide members of
Terzan~1, along with two additional RRab classified as dubious
members. These stars are listed in
Tables~\ref{tab_gcvar1}~and~\ref{tab_gcvar2}. All bona fide members
are located close to the cluster center ($r<1.8\arcmin\sim2\,R_h$).

The CVSGC for Terzan~1 includes nine RRab and three RRc. VIVACE fails to
detect three RRab and all RRc, likely due to their very central location
($r<1\arcmin$). Of the six RRab detected RRab in common, our analysis
classifies V9, V15 and V17 as bona fide members, V14 as a non-member,
and V7 and V10 as dubious members. Although the PM analysis indicates
a high probability of cluster membership for V7 and V10, their
distance moduli are $\sim0.2$ mag smaller than those of the other
RR~Lyrae stars.  This slight offset in their photometry may result
from their close proximity to the cluster center ($r<0.5\arcmin$).

\citet{pru24} agree with our membership assignments for all stars, and
consider our dubious members as bona fide members. In contrast,
\citet{cru24} disagree with our classification for V15 and V17, and
classify our dubious members as field stars instead. Additionally,
\citet{cru24} include one extra RRab member, Gaia-4055453741758171904,
which is also identified by VIVACE (326735) and is considered a bona
fide member in our analysis.

Finally, we highlight the identification of three new RRab and two new RRc
as bona fide cluster members. However, we note that \citet{cru24} also
identify one of the RRab, 326736, but classify it as a field star.

\subsection{Terzan~4 (HP~4)}
\label{sec_ter4}
The CVSGC does not list any variables for Terzan~4, and \citet{cru24}
do not include any RR~Lyrae stars for this GC either. However, one RRab
listed in VIVACE, 276896, is located relatively close to the cluster
center ($r\sim2.5\arcmin=1.7\,R_h$) and shows a high probability of
belonging to the cluster according to our PM analysis. We include it
as a new cluster variable candidate in
Tables~\ref{tab_gcvar1}~and~\ref{tab_gcvar2}.

\subsection{Terzan~5 (Terzan~11)}
\label{sec_ter5}
The CVSGC lists four RRab for Terzan~5, and \citet{cru24} add one extra
RRab, Gaia-4067312558805370624.  VIVACE detects only one of these,
V11=600303. Since our PM analysis assigns a high probability of
cluster membership to this star, and it lies in the inner regions of the GC
($r=1.2\arcmin\sim1.3\,R_h$), we classify it a probable cluster member
and include it in
Tables~\ref{tab_gcvar1}~and~\ref{tab_gcvar2}. \citet{pru24} also
classify it a cluster member. The only other inner-cluster RRab in our
analysis with high membership probability from its PM, and with a
similar distance modulus according to the PLZ relations (see
Fig.~\ref{fig_dis1}), is 600304, an RRc located at
$r=2.0\arcmin\sim2.2\,R_h$. We classify it as a new probable cluster
member but note \citet{cru24} consider it a field star.

\subsection{Terzan~9}
\label{sec_ter9}
The CVSGC lists no variables for Terzan~9, and \citet{cru24} do not
include any RR~Lyrae stars for this GC either. However, our XDGMM
algorithm assigns a high probability of cluster membership to the
VIVACE variable 805472, the RRc that is closest to the cluster center
($r=0.3\arcmin$). We report it as a probable candidate in
Tables~\ref{tab_gcvar1}~and~\ref{tab_gcvar2}.

\subsection{Terzan~10}
\label{sec_ter10}
The CVSGC lists ten RRab and one RRc as members of Terzan~10, based on
our \citetalias{alo15} and OGLE data \citep{sos14}. We reexamined the
variable population of this GC in \citetalias{alo21}, confirming the
RR~Lyrae members identified in \citetalias{alo15}. VIVACE recovers all
the RR~Lyrae members detected in both \citetalias{alo15} and
\citetalias{alo21} (seven RRab), plus one additional RRab, V51=860441, which
is included in the CVSGC from OGLE. All these stars are classified as
cluster members in our analysis and are listed in
Tables~\ref{tab_gcvar1}~and~\ref{tab_gcvar2}, along with 860314, an
RRc not previously listed in the CVSGC, which our analysis also
classifies as a bona fide cluster member. VIVACE also detects V12, an
RRab that is classified as a field star in the CVSGC. Our current
analysis similarly classifies it as a field star, and we flag it as
such in Table~\ref{tab_gcvar2}. \citet{pru24} agree with most of our
membership assignments but classify V12 and V51 differently, and they
are unable to classify V6 and V24. In contrast, \citet{cru24} classify
all variables stars listed in the CVSGC for Terzan~10 as field stars.

\subsection{UKS~1 (FSR~16)}
\label{sec_uks1}
We highlight the discovery of a significant RR~Lyrae variable
population in UKS~1.  Neither the CVSGC nor \citet{cru24} list any
variable stars for this GC. However, our analysis of the VIVACE
RR~Lyrae stars identifies nine RRab as bona fide cluster members, which
we have included in
Tables~\ref{tab_gcvar1}~and~\ref{tab_gcvar2}. While most of these
stars are concentrated near the cluster center
($r<1.9\arcmin\sim2.9\,R_h$), a few are located in the outskirts of
the cluster. Notably, their magnitudes are near the confusion limit,
which could account for the absence of a distinct HB in our CMD for
this GC (see Fig.~\ref{fig_cmd1}).

The average period of its RRab stars, 0.57 days (see
Table.~\ref{tab_gcvar2}), and its reported iron content, $[{\rm
    Fe/H}]=-1.0$ dex (see Table.~\ref{tab_gcinput}), classify UKS~1 as
a new member of the Oosterhoff I group \citep{cat15}.

\subsection{VVV-CL160}
\label{sec_vvv160}
We highlight the discovery of RR~Lyrae stars for VVV-CL160 in our
analysis.  This GC has not yet been included in the CVSGC, as it was only
recently confirmed as a new Milky Way GC by
\citet{min21c}. Although \citet{gar22a} examined it for RR~Lyrae stars using the OGLE database, no variables were found. Similarly, 
\citet{cru24} do not list any RR~Lyrae stars for this GC.
However, VIVACE lists several newly detected RR~Lyrae stars within the
tidal radius of VVV-CL160, and our PM analysis identifies the four RRab
closest to the GC center ($r<2.0\arcmin=1.3\,R_h$) as highly
probable members. The PM separation between cluster and
field stars for VVV-CL160 is particularly distinct (see
Sect.~\ref{sec_pm}), indicating a highly significant
membership distinction. Three of these RRab also have similar distance moduli
(see Fig.~\ref{fig_dis1}), and we classify as bona fide
cluster members in Tables~\ref{tab_gcvar1}~and~\ref{tab_gcvar2}. The
fourth RRab, 1055884, has a distance modulus derived from the PLZ
relations that is significantly smaller ($\Delta(K_s-M_{K_s})\sim0.5$ mag),
suggesting a shorter distance than the others, so we flagged it as a
dubious member.

\section{Distances and extinctions to the VVV GCs}
\label{sec_distance2}
We have used the near-infrared PLZ relations from
Sect.~\ref{sec_distance} along with the VIVACE periods and apparent
magnitudes of the RR~Lyrae stars in the VVV GCs in
Table~\ref{tab_gcvar2} to infer their apparent distance moduli and color
excesses (see Fig.~\ref{fig_dis1}). To obtain the true distance to
these RR~Lyrae stars, and therefore the distance to their parent GCs,
we need to correct for their color excesses using an extinction
law. Extinction at low latitudes toward the Galactic center varies
non-canonically \citep{nis06a,alo17,nog18,dek19}. Recently,
\citet{san22} recalibrated the infrared extinction law values in a
sizable region of $3\times3$ deg$^2$ around the Galactic center using
VVV, GLIMPSE and WISE data.  In our analysis we assume a \citet{car89}
canonical extinction law for the VVV GCs located farther from the
Galactic plane ($|l|>5^{\circ}$) and a \citet{san22} non-canonical
extinction law for those located at lower latitudes ($|l|>4^{\circ}$),
except for 2MASS-GC02, whose very severe differential extinctions (see
Fig.~\ref{fig_cmd1}) enable us to measure the selective-to-total
extinction ratios directly. As we did in \citetalias{alo15} and
\citetalias{alo21}, we perform an ordinary least-squares bisector fit
to the distance moduli and color excesses of the RR~Lyrae stars in
2MASS-GC02 inferred from the PLZ relations (see
Fig.~\ref{fig_dis1}). This results in selective-to-total extinction
ratios of $R_{K_s,H-K_s}=1.20\pm0.11$ and $R_{K_s,J-K_s}=0.56\pm0.06$
as the slopes of the linear fit. These values are in agreement with
those we derived in \citetalias{alo15} and \citetalias{alo21}.

The weighted average distances and color excesses of all the RR~Lyrae
stars in a given GC are provided in Table~\ref{tab_gcdis}, along with
their galactocentric distances, which we obtained assuming an
$R_0=8275\pm9\pm33$ pc \citep{gra21}. For GCs with more than one
RR~Lyrae star, only bona fide members were used to obtain their
parameters, excluding the RR~Lyrae stars flagged as dubious
candidates. As expected, the GCs located closer to the Galactic plane
exhibit higher reddenings. Even though at near-infrared wavelengths
extinction is diminished, color excesses increase from
$E(J-K_s)<0.2$~mag for GCs at $|l|>5^{\circ}$ up to values of
$E(J-K_s)\sim1.0$~mag for GCs closer to the Galactic plane, reaching
mean values as high as $E(J-K_s)=1.57$~mag for UKS~1,
$E(J-K_s)=1.79$~mag for VVV-CL~160, and even $E(J-K_s)=3.04$~mag for
2MASS-GC02, our most reddened GC.  Distances to the GCs place the vast
majority of them within the Galactic bulge, at distances from the
Galactic center $R_{GC}<3$~kpc (see Table~\ref{tab_gcdis}). The only
GCs in our sample outside the Galactic bulge realm are the nearby
NGC~6544 and NGC~6656, the highly reddened 2MASS-GC02, as well as the
distant UKS~1 and NGC~6441. Conversely, there are five GCs (FSR~1716,
NGC~6380, NGC~6401, NGC~6558, and NGC~6642) located in the innermost
regions of the bulge, at distances $R_{GC}<1$~kpc. 

\begin{figure}
  \centering
  \includegraphics[scale=0.75]{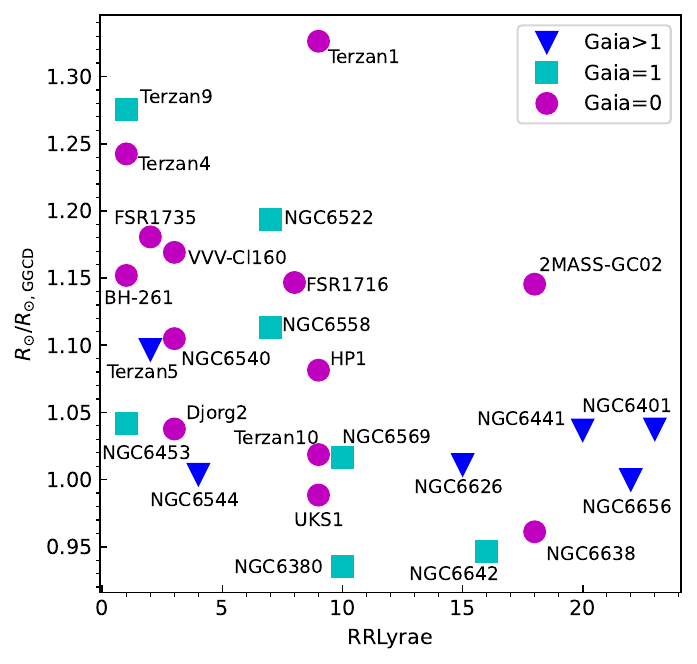}
  \caption{Ratio of distances calculated using RR~Lyrae stars in the VVV GCs compared to those from the GGCD, plotted against the number of RR~Lyrae stars present in each GCs. Blue inverted triangles represent GCs where \citet{bau21} were able to apply several techniques based on {\em Gaia} and HST data to measure distances, cyan squares indicate GCs where only one technique was used, while magenta dots show GCs lacking {\em Gaia} or HST measurements, relying solely on distances reported in the literature.}
  \label{fig_dis2}
\end{figure}

We compared the distances to the VVV GCs that we obtained from the
RR~Lyrae stars with those reported in the GGCD, which were taken
mainly from \citet{bau21}. They measured distances to the Galactic GCs
using different procedures based on {\em Gaia} or HST databases
and supplemented their measurements with distances provided in the
literature. Unfortunately, they could not apply all their procedures
to all of the Galactic GCs. In fact, we highlight that, out of the 26
VVV GCs with RR~Lyrae stars from our analysis, \citet{bau21} were able
to apply all their different procedures using {\em Gaia} and HST
data only for M~22, while for 12 GCs they apply at least one, and for the
remaining 13, the GGCD distance measurements come only from the
literature. As noted in Sect.~\ref{sec_distance}, we used M~22 to
recalibrate our near-infrared PLZ relations. In addition to being the
only cluster in our sample for which \citet{bau21} were able to apply
all their different procedures from the {\em Gaia} and HST data,
M~22 also contains the largest number of references in the
literature. For the remaining 25 VVV GCs with RR~Lyrae members for
which we derived distances, ten GCs show differences between our
distances and GGCD distances smaller than 5 percent, five GCs show
differences between 5 and 10 percent, and there are ten GCs for which
our RR~Lyrae-calculated distances are underestimated by more than 10
percent compared to those in the GGCD (see Fig.~\ref{fig_dis2}), with
differences as severe as $\sim33$ percent for Terzan~1.  We should
note, however, that this last group where the agreement is worst,
primarily contains GCs for which distances in the GGCD come from
literature. Conversely, most of the VVV GCs for which more than one
method based on {\em Gaia} or HST databases was applied in
\citet{bau21} are included in the group with smaller differences. We
should also highlight that most of the clusters for which only one
candidate RR~Lyrae star is detected are found in the group where the
agreement is poorer (see Fig.~\ref{fig_dis2}). The fact that they have
only one detected RR~Lyrae star, and that the latter are located in
their innermost, highly crowded regions ($r<1.75\,R_h$), renders their
derived distances and extinction values more prone to systematic
uncertainties affecting the RR~Lyrae photometry. An alternative, of
course, is that such stars do not, in fact, belong to the GCs.

\section{Conclusions}
\label{sec_conclusions}
We conducted a study of Galactic GCs located toward the innermost
regions of the Galaxy to search for their RR~Lyrae variable
stars. Using the recent VIVACE and VIRAC2 catalogs, both based on VVV
near-infrared observations, we explored the presence of these
classical pulsators in the 41 GCs listed in the GGCD within the VVV
survey footprint. To determine membership for each GC, we developed a
methodology that incorporates PM, projected proximity to the cluster
center, and distance moduli derived from the PLZ relations. As a
by-product of this analysis, we also calculated the PM for all these
VVV GCs, independently confirming {\em Gaia}'s measurements except for
two of the most reddened GCs: UKS~1 and 2MASS-GC02.

Our study found RR~Lyrae variable stars in 26 of the sampled VVV
GCs. The only clusters lacking RR~Lyrae stars were either metal-rich
GCs or a few sparsely populated metal-poor GCs. We successfully
identified most of the known RR~Lyrae stars listed in the CVSGC for
these GCs, missing only those located in the crowded central regions,
where even PSF photometry struggles to resolve individual stellar
sources. On the other hand, we were able to assign membership to
RR~Lyrae stars located in the outskirts of the GCs, which had not been
previously considered. Our membership assignments largely align with
those from recent studies in the literature for most of the RR~Lyrae
stars. Notably, our analysis led to the discovery of unreported
RR~Lyrae stars in 16 VVV GCs, including the first detection of several
of these classic pulsators in the highly reddened clusters UKS~1 and
VVV-CL160. We also identified one RR~Lyrae star as a potential
candidate in each of Terzan~4 and Terzan~9.

Building on the tight PLZ relations exhibited by RR~Lyrae stars in the
near-infrared, and assuming a \citet{car89} canonical extinction law
for those VVV GCs with $|l|>5^{\circ}$ and a \citet{san22}
non-canonical law for those VVV GCs closer to the Galactic plane, with
$|l|<4^{\circ}$, we provided distances and near-infrared
color-excesses for 26 low-latitude GCs. As expected, the GCs closer to
the Galactic plane were found to be the most heavily reddened. Our
analysis also revealed that all but five of these GCs are located
within the Galactic bulge. Five of our sampled GCs --FSR~1716,
NGC~6380, NGC~6401, NGC~6558, and NGC~6642-- are even located less
than 1~kpc from the Galactic center.

We compared our RR~Lyrae-derived distances with those from the
GGCD. We found good agreement, within the error, and typically within
5 percent, for the GCs for which the GGCD provides distances measured
based on {\em Gaia}. However, there are ten GCs whose distances differ
significantly, mostly encompassing GCs with no {\em Gaia}
measurement. The averaged distances from the literature provided by
the GGCD for the GCs in this group is usually underestimated by more
than 10 percent. Most of them contain a significant number of RR~Lyrae
members, which provide compatible distances to support our results,
even though we recommend some caution in those few cases where a
cluster contains a single identified RR~Lyrae member.

\begin{acknowledgements}
J.A.-G., M.C., C.E.F.L, and Z.G. are supported by ANID's Millennium
Science Initiative through grants ICN12\textunderscore009 and
AIM23-0001, awarded to the Millennium Institute of Astrophysics (MAS).
J.A.-G. also acknowledges support from ANID's FONDECYT Regular grants
\#1201490. M.C. and C.E.F.L also acknowledges support from ANID's
FONDECYT Regular grants \#1201490 and \#1231637. Support for M.C. is
also provided by ANID's Basal project FB210003.  C.E.F.L is also
supported by DIUDA 88231R11; by LSST Discovery Alliance grant; by
GEMINI grant 32240028. J.G.F-T gratefully acknowledges the grants
support provided by ANID's FONDECYT Iniciaci\'on No. 11220340, ANID
Fondecyt Postdoc No. 3230001 (Sponsoring researcher), from the Joint
Committee ESO-Government of Chile under the agreement 2021 ORP
023/2021 and 2023 ORP 062/2023. P.W.L. acknowledges support by STFC
grant ST/Y000846/1. R.K.S. acknowledges support from CNPq/Brazil
through projects 308298/2022-5 and 421034/2023-8. Z.G. acknowledges
the support from ANID's FONDECYT postdoctoral programme 3220029.  The
authors gratefully acknowledge the use of data from the ESO Public
Survey program ID 179.B-2002 and 198.B2004, taken with the VISTA
telescope, and data products from the Cambridge Astronomical Survey
Unit.
\end{acknowledgements}

\bibliographystyle{aa} 
\bibliography{mybibtex} 

\begin{appendix}
  \begin{table*}
  \section{Positions and physical parameters of the VVV GCs}
  \caption{Positions and physical parameters of the target clusters.}
  \label{tab_gcinput}
  \centering
  \begin{tabular}{cccccccccc}
    \hline\hline
    Cluster & $\alpha$(J2000) & $\delta$(J2000) & $l$ & $b$ & $[{\rm Fe/H}]$ & $M_V$ & Mass & $R_{h}$ & $R_{t}$ \\
    & (h:m:s) & (d:m:s) & (deg) & (deg) & (dex) & (mag) & ($\times10^4M_{\odot}$) & (arcmin) & (arcmin) \\
    \hline
  2MASS-GC02 & 18:09:36.50 & -20:46:44.0 & 9.782   & -0.615  & -1.08$^a$ &  24.6 &  1.56  & 1.39 & 9.81 \\
  BH~261     & 18:14:06.60 & -28:38:06.0 & 3.362   & -5.27   & -1.1 $^b$ &  11.26&  2.38  & 1.65 & 11.59\\
  Djorg~1     & 17:47:28.70 & -33:03:59.0 & 356.675 & -2.484  & -1.36$^c$ &  13.08&  8.43  & 1.43 & 8.82 \\
  Djorg~2     & 18:01:49.05 & -27:49:32.9 & 2.763   & -2.508  & -1.07$^d$ &  10.7 &  13.4  & 1.82 & 6.06 \\
  FSR~1716   & 16:10:30.00 & -53:44:56.0 & 329.778 & -1.593  & -1.38$^e$ &  13.06&  5.66  & 1.71 & 15.79\\
  FSR~1735   & 16:52:10.60 & -47:03:29.0 & 339.188 & -1.853  & -0.9 $^f$ &  14.38&  9.98  & 0.79 & 13.37\\
  Gran~1     & 17:58:36.24 & -32:01:12.0 & 358.767 & -3.977  & -1.19$^g$ &  12.4 &  2.61  & 2.38 & 7.92 \\
  Gran~5     & 17:48:54.72 & -24:10:12.0 & 4.459   & 1.839   & -1.56$^g$ &  12.11&  2.29  & 1.21 & 15.06\\
  HP~1       & 17:31:05.20 & -29:58:54.0 & 357.425 & 2.115   & -1.21$^d$ &  11.07&  13.7  & 1.52 & 9.11 \\
  Liller~1   & 17:33:24.56 & -33:23:22.4 & 354.84  & -0.161  & -0.14$^d$ &  15.73&  101.0 & 0.55 & 16.58\\
  NGC~6380   & 17:34:28.47 & -39:04:10.3 & 350.182 & -3.422  & -0.78$^d$ &  10.7 &  34.1  & 1.12 & 14.39\\
  NGC~6401   & 17:38:36.53 & -23:54:34.6 & 3.45    & 3.98    & -1.09$^d$ &  9.97 &  12.1  & 1.07 & 7.31 \\
  NGC~6440   & 17:48:52.84 & -20:21:37.5 & 7.729   & 3.801   & -0.53$^h$ &  8.97 &  56.9  & 0.55 & 14.12\\
  NGC~6441   & 17:50:13.06 & -37:03:05.2 & 353.532 & -5.006  & -0.49$^d$ &  7.12 &  139.0 & 0.58 & 29.25\\
  NGC~6453   & 17:50:51.72 & -34:35:54.5 & 355.718 & -3.872  & -1.57$^c$ &  9.19 &  16.8  & 0.94 & 11.24\\
  NGC~6522   & 18:03:34.07 & -30:02:02.3 & 1.025   & -3.926  & -1.22$^d$ &  8.28 &  21.4  & 1.17 & 7.24 \\
  NGC~6528   & 18:04:49.61 & -30:03:20.8 & 1.139   & -4.174  & -0.16$^d$ &  9.71 &  9.44  & 1.08 & 4.82 \\
  NGC~6540   & 18:06:08.56 & -27:45:55.0 & 3.285   & -3.313  & -1.02$^d$ &  9.74 &  5.62  & 1.54 & 13.21\\
  NGC~6544   & 18:07:20.12 & -24:59:53.6 & 5.837   & -2.202  & -1.52$^d$ &  7.86 &  8.15  & 2.05 & 63.62\\
  NGC~6553   & 18:09:17.52 & -25:54:29.0 & 5.253   & -3.029  & -0.19$^d$ &  8.04 &  22.9  & 1.48 & 30.85\\
  NGC~6558   & 18:10:17.75 & -31:45:52.2 & 0.199   & -6.023  & -0.99$^d$ &  9.66 &  3.13  & 0.68 & 3.95 \\
  NGC~6569   & 18:13:38.80 & -31:49:36.8 & 0.481   & -6.681  & -0.92$^d$ &  8.9  &  22.9  & 0.85 & 14.01\\
  NGC~6624   & 18:23:40.51 & -30:21:39.7 & 2.788   & -7.913  & -0.69$^i$ &  8.03 &  10.3  & 0.73 & 7.21 \\
  NGC~6626   & 18:24:32.89 & -24:52:11.4 & 7.798   & -5.581  & -1.29$^j$ &  6.85 &  27.0  & 1.03 & 30.36\\
  NGC~6637   & 18:31:23.10 & -32:20:53.1 & 1.723   & -10.269 & -0.59$^k$ &  7.62 &  13.8  & 0.93 & 9.35 \\
  NGC~6638   & 18:30:56.10 & -25:29:50.9 & 7.896   & -7.153  & -0.99$^k$ &  8.79 &  12.4  & 0.65 & 10.21\\
  NGC~6642   & 18:31:54.23 & -23:28:32.2 & 9.815   & -6.439  & -1.09$^d$ &  9.65 &  3.95  & 0.59 & 6.93 \\
  NGC~6656   & 18:36:23.94 & -23:54:17.1 & 9.892   & -7.552  & -1.7 $^d$ &  5.06 &  47.0  & 3.31 & 80.03\\
  Pal~6      & 17:43:42.19 & -26:13:30.0 & 2.09    & 1.779   & -0.92$^d$ &  11.6 &  8.56  & 1.11 & 5.09 \\
  Terzan~1      & 17:35:47.20 & -30:28:54.4 & 357.558 & 0.991   & -1.26$^l$ &  12.4 &  19.9  & 0.89 & 23.17\\
  Terzan~2      & 17:27:33.10 & -30:48:08.4 & 356.319 & 2.298   & -0.86$^d$ &  13.11&  8.05  & 1.06 & 5.52 \\
  Terzan~4      & 17:30:39.00 & -31:35:43.9 & 356.024 & 1.308   & -1.38$^d$ &  13.43&  18.1  & 1.48 & 7.11 \\
  Terzan~5      & 17:48:04.85 & -24:46:44.6 & 3.839   & 1.687   & -0.78$^d$ &  12.36&  109.0 & 0.92 & 26.62\\
  Terzan~6      & 17:50:46.38 & -31:16:31.4 & 358.571 & -2.162  & -0.65$^m$ &  14.47&  10.0  & 0.5  & 6.57 \\
  Terzan~9      & 18:01:38.80 & -26:50:23.0 & 3.603   & -1.989  & -1.36$^d$ &  12.73&  13.7  & 0.99 & 18.83\\
  Terzan~10     & 18:02:57.80 & -26:04:01.0 & 4.421   & -1.864  & -1.62$^d$ &  14.73&  31.9  & 1.16 & 13.99\\
  Terzan~12     & 18:12:15.80 & -22:44:31.0 & 8.358   & -2.101  & -0.56$^d$ &  13.82&  3.76  & 1.18 & 16.55\\
  Ton~2      & 17:36:10.08 & -38:33:22.0 & 350.793 & -3.424  & -0.74$^d$ &  11.66&  4.31  & 1.41 & 10.6 \\
  UKS~1      & 17:54:27.19 & -24:08:43.0 & 5.125   & 0.764   & -1.0 $^d$ &  17.3 &  7.99  & 0.66 & 10.6 \\
  VVV-CL001  & 17:54:42.50 & -24:00:53.0 & 5.267   & 0.78    & -2.45$^n$ &  14.2 &  15.4  & 1.0  & 15.5 \\
  VVV-CL160  & 18:06:57.00 & -20:00:40.0 & 10.151  & 0.302   & -1.32$^o$ &  15.6 &  5.28  & 1.66 & 9.87 \\
    \hline
  \end{tabular}
  \tablefoot{Equatorial and Galactic coordinates, absolute integrated visual magnitudes, masses, half-light and tidal radii are taken from the GGCD. Iron contents are taken from $^a$\citet{pen15}; $^b$\citet{kun24}; $^c$\citet{vas18}; $^d$\citet{sch24}; $^e$\citet{koc17}; $^f$\citet{car16}; $^g$\citet{gra22}; $^h$\citet{cro23}; $^i$\citet{val11b};  $^j$\citet{vil17}; $^k$\citet{car09}; $^l$\citet{val15}; $^m$\citet{fan24}; $^n$\citet{fer21}; $^o$Garro et al. (in prep).}
\end{table*}

  \begin{figure*}
  \section{Near-infrared CMDs for the VVV GCs}
    \centering
    \includegraphics[scale=1.0]{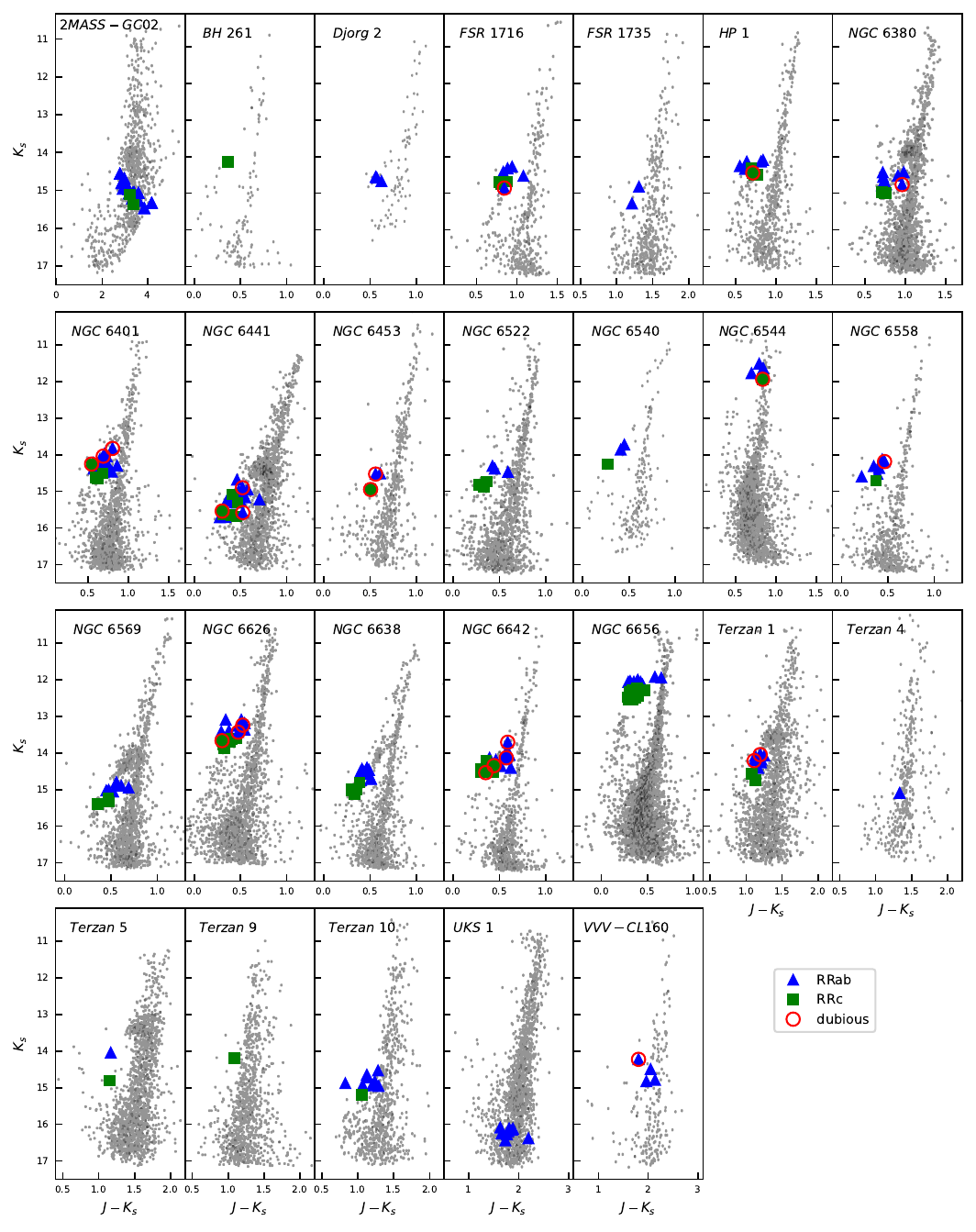}
    \caption{Near-infrared CMDs for the VVV GCs with assigned RR~Lyrae
    members. The diagrams include only stars located in the innermost
    GC regions ($r\leq1.0\arcmin$), selected as candidates by our PM
    XDGMM classifier. RR~Lyrae stars for each GC are overplotted, with
    RRab stars represented as blue triangles and RRc stars as green
    squares. Dubious candidates are highlighted with red circles.}
    \label{fig_cmd1}
  \end{figure*}

  \begin{figure*}
    \centering
    \includegraphics[scale=1.0]{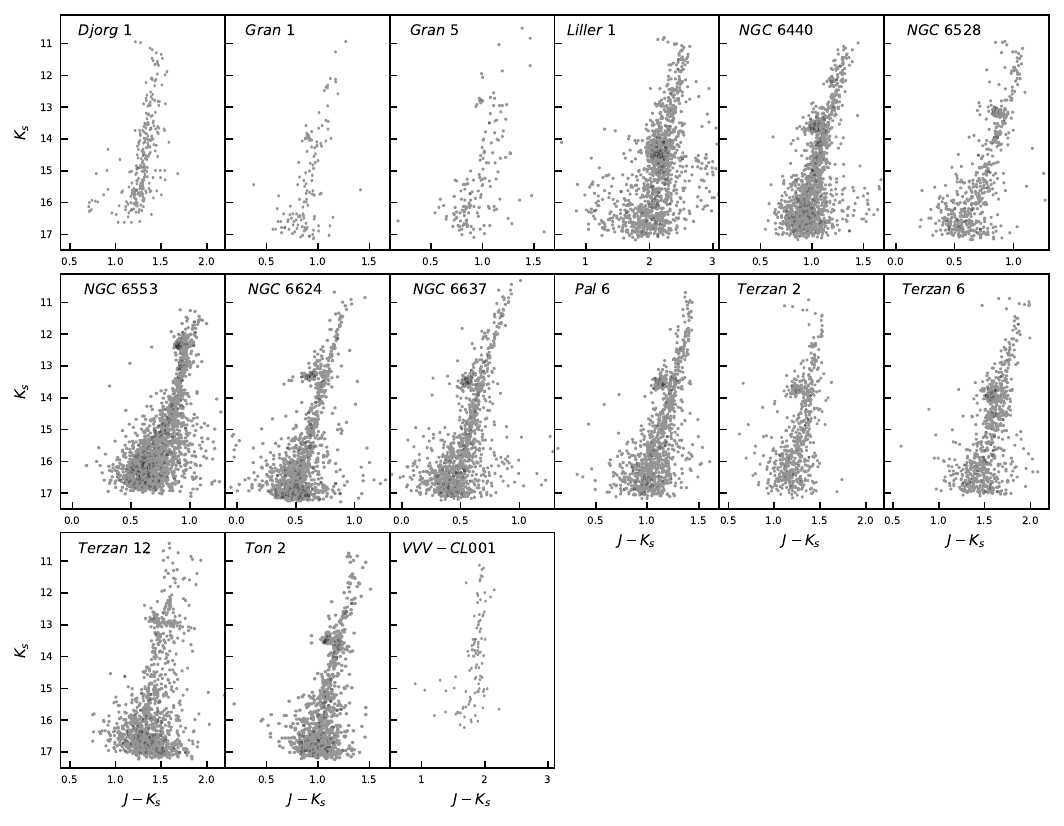}
    \caption{Near-infrared CMDs for the VVV GCs with no RR~Lyrae
      members. To reduce contamination, only stars located within
      $r\leq1.0\arcmin$ of the cluster centers and selected as
      candidates by our PM XDGMM classifier are included in the
      plots.}
    \label{fig_cmd2}
  \end{figure*}

  \begin{figure*}
  \section{VPDs and PM for the VVV GCs}
    \centering
    \includegraphics[scale=1.0]{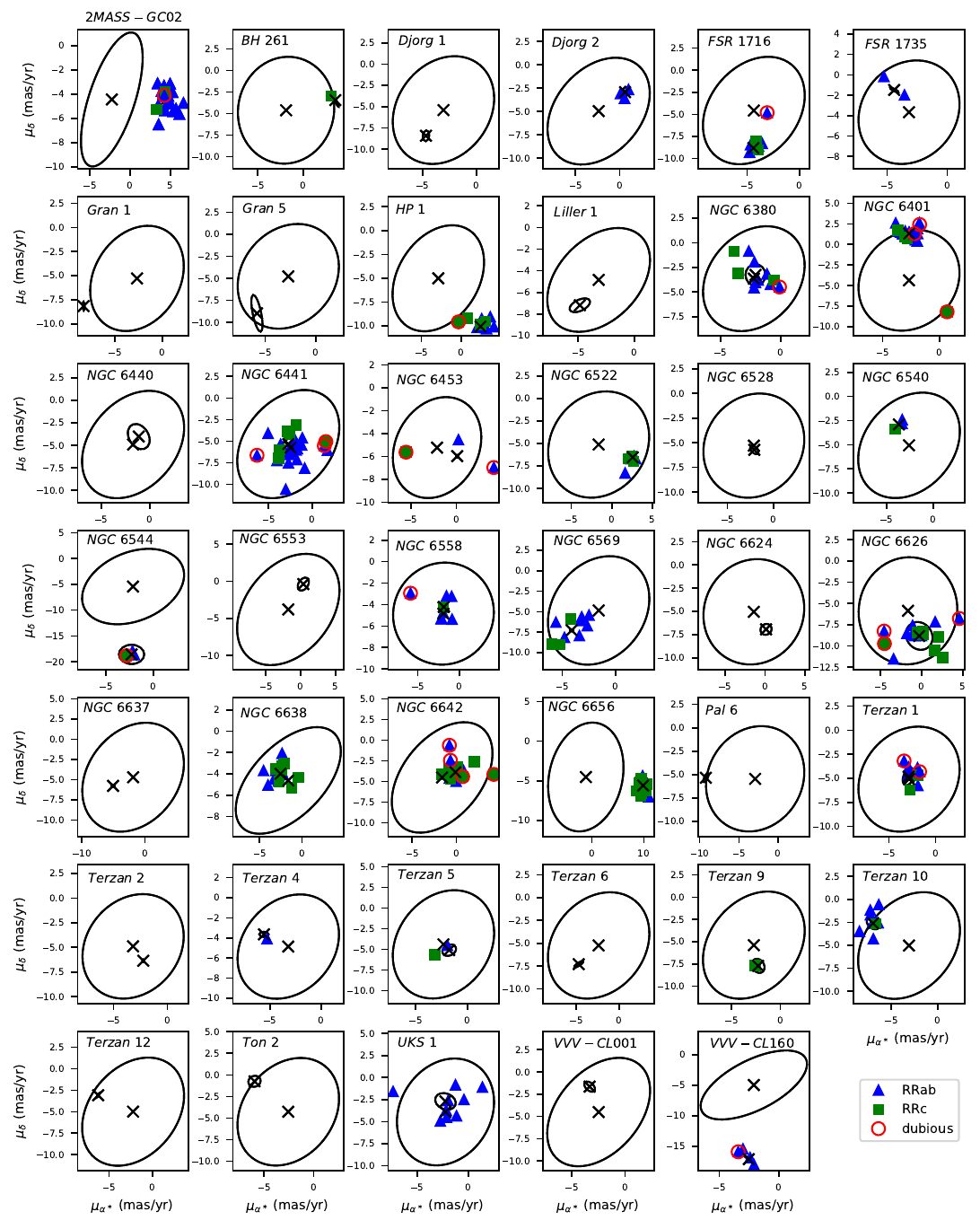}
    \caption{VPDs showing the PM of the RR~Lyrae cluster members. RRab stars are represented as blue triangles, while RRc stars are shown as green squares. Red circles indicate dubious candidates. Ellipses representing the $2\sigma$ distributions of field and cluster stars, as determined by our XDGMM algorithm, are overplotted, with crosses marking the centers of both distributions.}
    \label{fig_pm}
  \end{figure*}

  \begin{table*}
  \caption{Proper motions of the VVV GCs.} 
  \label{tab_gcpm}
  \centering
  \begin{tabular}{ccccc}
    \hline\hline
    Cluster & $\mu_{\alpha^\ast}$ & $\mu_{\delta}$ & $\mu_{\alpha^\ast GGCD}$ & $\mu_{\delta GGCD}$ \\
     & (mas yr$^{-1}$) & (mas yr$^{-1}$) & (mas yr$^{-1}$) & (mas yr$^{-1}$) \\
    \hline
  2MASS-GC~02& $4.2\pm0.4$ & $-4.6\pm0.4$ & $-2.924\pm0.22 $ & $-3.776\pm0.174$ \\  
  BH~261     & $3.6\pm0.2$ & $-3.5\pm0.4$ & $3.589\pm0.022$ & $-3.57\pm0.02 $ \\  
  Djorg~1    & $-4.7\pm0.1$ & $-8.4\pm0.3$ & $-4.725\pm0.028$ & $-8.47\pm0.021$ \\  
  Djorg~2    & $0.6\pm0.2$ & $-2.9\pm0.3$ & $0.72\pm0.021$ & $-3.003\pm0.018$ \\  
  FSR~1716   & $-4.40\pm0.24$ & $-8.82\pm0.24$ & $-4.407\pm0.029$ & $-8.862\pm0.026$ \\  
  FSR~1735   & $-4.41\pm0.18$ & $-1.49\pm0.11$ & $-4.57\pm0.054$ & $-1.515\pm0.046$ \\  
  Gran~1     & $-8.27\pm0.05$ & $-8.16\pm0.22$ & $-8.044\pm0.038$ & $-8.049\pm0.031$ \\  
  Gran~5     & $-5.8\pm0.3$ & $-9.0\pm1.0$ & $-5.353\pm0.046$ & $-9.312\pm0.043$ \\  
  HP~1       & $2.43\pm0.11$ & $-10.07\pm0.22$ & $2.5\pm0.028$ & $-10.04\pm0.027$ \\  
  Liller~1   & $-4.8\pm0.5$ & $-7.2\pm0.3$ & $-5.183\pm0.176$ & $-6.979\pm0.143$ \\  
  NGC~6380   & $-2.1\pm0.4$ & $-3.3\pm0.5$ & $-2.186\pm0.017$ & $-3.256\pm0.016$ \\  
  NGC~6401   & $-2.68\pm0.18$ & $1.32\pm0.10$ & $-2.762\pm0.01 $ & $1.464\pm0.008$ \\  
  NGC~6440   & $-1.2\pm0.6$ & $-4.0\pm0.7$ & $-1.189\pm0.024$ & $-4.004\pm0.024$ \\  
  NGC~6441   & $-2.5\pm0.5$ & $-5.8\pm0.5$ & $-2.566\pm0.02 $ & $-5.373\pm0.02 $ \\  
  NGC~6453   & $0.10\pm0.07$ & $-5.98\pm0.07$ & $0.165 \pm0.021$ & $-5.92 \pm0.021$ \\  
  NGC~6522   & $2.55\pm0.20$ & $-6.55\pm0.05$ & $2.586 \pm0.014$ & $-6.432\pm0.014$ \\  
  NGC~6528   & $-2.10\pm0.24$ & $-5.72\pm0.23$ & $-2.173\pm0.021$ & $-5.643\pm0.019$ \\  
  NGC~6540   & $-3.68\pm0.04$ & $-2.89\pm0.10$ & $-3.695\pm0.019$ & $-2.818\pm0.018$ \\  
  NGC~6544   & $-2.3\pm0.7$ & $-18.6\pm0.9$ & $-2.269\pm0.026$ & $-18.61\pm0.026$ \\  
  NGC~6553   & $0.4\pm0.4$ & $-0.4\pm0.4$ & $0.354\pm0.013$ & $-0.427\pm0.013$ \\  
  NGC~6558   & $-1.87\pm0.25$ & $-4.22\pm0.13$ & $-1.752\pm0.017$ & $-4.167\pm0.016$ \\  
  NGC~6569   & $-4.10\pm0.08$ & $-7.27\pm0.07$ & $-4.142\pm0.01 $ & $-7.333\pm0.009$ \\  
  NGC~6624   & $0.1\pm0.3$ & $-6.9\pm0.3$ & $0.127 \pm0.014$ & $-6.956\pm0.014$ \\  
  NGC~6626   & $-0.3\pm0.8$ & $-8.8\pm0.8$ & $-0.303\pm0.018$ & $-8.931\pm0.018$ \\  
  NGC~6637   & $-5.05\pm0.13$ & $-5.76\pm0.08$ & $-5.053\pm0.007$ & $-5.825\pm0.007$ \\  
  NGC~6638   & $-2.54\pm0.12$ & $-4.00\pm0.06$ & $-2.515\pm0.011$ & $-4.069\pm0.01 $ \\  
  NGC~6642   & $-0.1\pm0.3$ & $-3.9\pm0.2$ & $-0.175\pm0.013$ & $-3.9  \pm0.013$ \\  
  NGC~6656   & $9.8\pm0.6$ & $-5.6\pm0.6$ & $9.841\pm0.009$ & $-5.607\pm0.009$ \\  
  Pal~6      & $-9.20\pm0.17$ & $-5.34\pm0.25$ & $-9.19\pm0.021$ & $-5.283\pm0.018$ \\  
  Terzan~1   & $-2.9\pm0.3$ & $-4.8\pm0.4$ & $-2.886\pm0.046$ & $-4.87\pm0.044$ \\  
  Terzan~2   & $-2.24\pm0.04$ & $-6.35\pm0.07$ & $-2.142\pm0.03 $ & $-6.255\pm0.©027$ \\  
  Terzan~4   & $-5.58\pm0.23$ & $-3.64\pm0.13$ & $-5.364\pm0.035$ & $-3.674\pm0.031$ \\  
  Terzan~5   & $-1.8\pm0.3$ & $-5.1\pm0.3$ & $-1.864\pm0.03 $ & $-5.108\pm0.027$ \\  
  Terzan~6   & $-4.7\pm0.3$ & $-7.3\pm0.2$ & $-4.993\pm0.052$ & $-7.458\pm0.045$ \\  
  Terzan~9   & $-2.3\pm0.3$ & $-7.7\pm0.4$ & $-2.168\pm0.031$ & $-7.698\pm0.029$ \\  
  Terzan~10  & $-6.9\pm0.3$ & $-2.6\pm0.3$ & $-6.806\pm0.03 $ & $-2.595\pm0.023$ \\  
  Terzan~12  & $-6.24\pm0.08$ & $-3.08\pm0.13$ & $-6.234\pm0.016$ & $-3.063\pm0.013$ \\  
  Ton~2      & $-6.0\pm0.3$ & $-0.7\pm0.3$ & $-5.936\pm0.02 $ & $-0.795\pm0.019$ \\  
  UKS~1      & $-2.2\pm0.5$ & $-2.7\pm0.5$ & $-0.96 \pm0.207$ & $-2.941\pm0.135$ \\  
  VVV-CL~001 & $-3.3\pm0.3$ & $-1.6\pm0.3$ & $-3.394\pm0.11 $ & $-1.71 \pm0.077$ \\  
  VVV-CL~160 & $-2.5\pm0.3$ & $-17.1\pm0.4$ & $-2.025\pm0.133$ & $-16.76\pm0.104$ \\  
    \hline
  \end{tabular}
\end{table*}

\begin{table*}
  \section{RR~Lyrae stars associated with the VVV GCs}
  \caption{Number of RR~Lyrae stars associated with the VVV GCs.}
  \label{tab_gcvar1}
  \centering
  \begin{tabular}{ccccccccc}
    \hline\hline
    Cluster & \multicolumn{2}{c}{CVSGC\tablefootmark{a}} & \multicolumn{2}{c}{Recent literature adds\tablefootmark{b}} & \multicolumn{2}{c}{This work\tablefootmark{c}} & \multicolumn{2}{c}{New in this work} \\
     & RRab & RRc & RRab & RRc & RRab & RRc & RRab & RRc \\
    \hline
    2MASS-GC02 & 13 & 0 & 2 & 0 & 16(+1) & 2 & 2(+1) & 2 \\
    BH~261 (AL~3) & -- & -- & 0 & 1 & 0 & (1) & 0 & 0 \\
    Djorg~2 & 5 & 2 & -- & -- & 3 & 0 & 0 & 0 \\
    FSR~1716 (VVV-GC05) & -- & -- &  6 & 3 & 5 & 3 & 0 & 0 \\
    FSR~1735 (2MASS-GC03) & -- & -- & 3 & 0 & 2 & 0 & 0 & 0 \\
    HP~1 (FSR~1781) & 0 & 0 & 8 & 5 & 6 & 3(+1) & 0 & 0 \\
    NGC~6380 (Ton~1) & 0 & 0 & 7 & 3 & 7(+1) & 3 & 1(+1) & 0 \\
    NGC~6401 & 23 & 11 & 7 & -- & 21(+2) & 3(+1) & 0 & 0 \\
    NGC~6441 & 47 & 26 & 2 & 1 & 20(+2) & 5(+1) & 2(+1) & 1 \\
    NGC~6453 & 3 & 5 & -- & -- & 1(+1)& (1) & 0 & 0 \\
    NGC~6522 & 5 & 6 & 0 & 1 & 4 & 3 & 1 & 0 \\
    NGC~6540 (Djorg~3)& 2 & 1 & -- & -- & 2 & 1 & 1 & 0 \\
    NGC~6544 & 1 & 0 & 1 & -- & 3 & (1) & 1 & (1) \\
    NGC~6558 & 7 & 3 & -- & -- & 6(+1) & 1 & (1) & 0 \\    
    NGC~6569 & 12 & 12 & 2 & 0 & 7 & 3 & 0 & 0 \\
    NGC~6626 (M~28) & 10 & 8 & 1 & 3 & 8(+2) & 6(+1) & 0 & 0 \\
    NGC~6638 & 10 & 16 & 3 & 0 & 10 & 7 & 1 & 0 \\
    NGC~6642 & 11 & 6 & 1 & 2 & 8(+2) & 8(+2) & 1 & 1(+2) \\
    NGC~6656 (M~22) & 11 & 15 & 0 & 0 & 10 & 13 & 0 & 0 \\
    Terzan~1 (HP~2)& 9 & 3 & 1 & -- & 7(+2) & 2 & 3 & 2 \\
    Terzan~4 (HP~4)& 0 & 0 & -- & -- & (1) & 0 & (1) & 0 \\
    Terzan~5 (Terzan~11)& 4 & 0 & 1 & 0 & (1) & (1) & 0 & (1) \\
    Terzan~9 & 0 & 0 & -- & -- & 0 & (1) & 0 & (1) \\
    Terzan~10 & 10 & 1 & -- & -- & 8 & 1 & 0 & 1 \\
    UKS~1 (FSR~16) & 0 & 0 & -- & -- & 9 & 0 & 9 & 0 \\
    VVV-CL160 & -- & -- & 0 & 0 & 3(+1) & 0 & 3(+1) & 0 \\
  \end{tabular}
  \tablefoot{
    \tablefoottext{a}{RR~Lyrae stars whose variability type is only suspected were included, while RR~Lyrae flagged as field stars in the CVSGC were discarded.}
  \tablefoottext{b}{References are available in the text from the different subsections in Sect.~\ref{sec_var}}
  \tablefoottext{c}{In parenthesis, we include those VIVACE RR~Lyrae detections which cluster membership is dubious, as detailed in the different subsections in Sect.~\ref{sec_var}}
  }
\end{table*}

\begin{table*}
  \caption{Observational parameters of the RR~Lyrae stars associated with the VVV GCs.}
  \label{tab_gcvar2}
  \centering
  \resizebox{\textwidth}{!}{
  \begin{tabular}{cccccccccccccccccc}
    \hline\hline
    Cluster  & VIVACE\_ID & CVSGC\_ID & VIRAC2\_ID & $\alpha$(J2000) & $\delta$(J2000) & Distance & Period & $K_s$\_amp & Z & Y & J & H & $K_s$ & $\mu_{\alpha^\ast}$ & $\mu_{\delta}$ & Type & Flag\tablefootmark{a} \\
    & & & & (h:m:s) & (d:m:s) & (arcmin) & (days) & (mag) & (mag) & (mag) & (mag) & (mag) & (mag) & (mas yr$^{-1}$) & (mas yr$^{-1}$) & &  \\
    \hline
terzan10 & 860249 & 24 & 13251592020129 & 18:02:46.96 & -26:06:09.1 & 3.24 & 0.592177 & 0.260 & 17.509$\pm$0.016 & 16.574$\pm$0.016 & 16.144$\pm$0.007 & 15.293$\pm$0.010 & 14.945$\pm$0.002 & -7.23$\pm$0.60 & -1.17$\pm$0.58 & RRab & 0 \\  
terzan10 & 860296 & 7  & 13251592002461 & 18:02:53.24 & -26:05:12.7 & 1.57 & 0.715332 & 0.569 & 17.557$\pm$0.022 & 16.621$\pm$0.018 & 15.847$\pm$0.008 & 15.104$\pm$0.018 & 14.724$\pm$0.002 & -8.40$\pm$0.80 & -3.49$\pm$0.84 & RRab & 0 \\  
terzan10 & 860314 & -- & 13255688010724 & 18:02:56.20 & -26:05:37.4 & 1.65 & 0.363141 & 0.204 & 18.010$\pm$0.017 & 17.026$\pm$0.017 & 16.258$\pm$0.005 & 15.528$\pm$0.007 & 15.202$\pm$0.002 & -6.66$\pm$0.59 & -2.64$\pm$0.57 & RRc  & 0 \\  
terzan10 & 860319 & 22 & 13255688015979 & 18:03:01.41 & -26:06:48.0 & 2.90 & 0.618602 & 0.519 & 17.266$\pm$0.014 & 16.556$\pm$0.014 & 15.710$\pm$0.006 & 15.072$\pm$0.010 & 14.882$\pm$0.001 & -6.99$\pm$0.46 & -2.60$\pm$0.44 & RRab & 0 \\  
terzan10 & 860326 & 12 & 13255688011890 & 18:03:04.08 & -26:05:33.7 & 2.09 & 0.568659 & 0.377 & 17.150$\pm$0.015 & 16.277$\pm$0.011 & 15.531$\pm$0.005 & 14.858$\pm$0.008 & 14.533$\pm$0.001 & -1.42$\pm$0.41 & -7.33$\pm$0.40 & RRab & 2 \\  
terzan10 & 860332 & 6  & 13255688015607 & 18:02:56.90 & -26:05:19.7 & 1.33 & 0.582333 & 0.334 & 18.323$\pm$0.021 & 17.008$\pm$0.020 & 16.009$\pm$0.005 & 15.282$\pm$0.007 & 14.940$\pm$0.001 & -6.38$\pm$0.49 & -2.57$\pm$0.47 & RRab & 0 \\  
terzan10 & 860342 & 2  & 13255688014228 & 18:02:59.48 & -26:04:22.6 & 0.52 & 0.730556 & 0.426 & 17.491$\pm$0.026 & 16.442$\pm$0.014 & 15.779$\pm$0.008 & 14.993$\pm$0.013 & 14.651$\pm$0.002 & -6.91$\pm$1.17 & -4.29$\pm$1.15 & RRab & 0 \\  
terzan10 & 860372 & 3  & 13251592013442 & 18:02:54.04 & -26:03:46.9 & 0.88 & 0.701721 & 0.241 & 18.132$\pm$0.023 & 16.880$\pm$0.021 & 16.072$\pm$0.005 & 15.316$\pm$0.007 & 14.834$\pm$0.002 & -7.22$\pm$0.53 & -1.68$\pm$0.54 & RRab & 0 \\  
terzan10 & 860438 & 5  & 13247496010643 & 18:02:57.14 & -26:02:44.0 & 1.29 & 0.688556 & 0.317 & 18.511$\pm$0.026 & 17.348$\pm$0.034 & 16.251$\pm$0.007 & 15.296$\pm$0.008 & 14.959$\pm$0.002 & -6.61$\pm$0.60 & -2.46$\pm$0.61 & RRab & 0 \\  
terzan10 & 860441 & 51 & 13247496003263 & 18:02:59.07 & -26:02:36.5 & 1.44 & 0.755820 & 0.308 & 17.532$\pm$0.024 & 16.606$\pm$0.030 & 15.820$\pm$0.010 & 15.061$\pm$0.016 & 14.538$\pm$0.002 & -6.32$\pm$0.83 & -0.57$\pm$0.86 & RRab & 0 \\  
    \hline
uks1     & 617101 & -- & 13071343008065 & 17:54:02.93 & -24:16:57.9 & 9.93 & 0.564883 & 0.412 &       --      & 18.961$\pm$0.066 & 18.008$\pm$0.018 & 16.846$\pm$0.021 & 16.124$\pm$0.006 & -0.43$\pm$1.76 & -2.50$\pm$1.72 & RRab & 0 \\  
uks1     & 617141 & -- & 13071343010726 & 17:54:05.26 & -24:15:43.0 & 8.60 & 0.578788 & 0.348 &       --      & 19.121$\pm$0.097 & 17.989$\pm$0.023 & 16.713$\pm$0.023 & 16.197$\pm$0.008 & -1.26$\pm$2.16 & -0.84$\pm$2.19 & RRab & 0 \\  
uks1     & 617430 & -- & 13063152004586 & 17:54:22.93 & -24:09:56.7 & 1.57 & 0.504499 & 0.453 &       --      & 19.179$\pm$0.168 & 18.175$\pm$0.025 & 17.069$\pm$0.030 & 16.443$\pm$0.010 & -2.13$\pm$2.96 & -4.52$\pm$2.83 & RRab & 0 \\  
uks1     & 617434 & -- & 13063152000147 & 17:54:22.95 & -24:09:21.2 & 1.16 & 0.515499 & 0.351 &       --      & 19.231$\pm$0.104 & 18.050$\pm$0.024 & 16.859$\pm$0.027 & 16.284$\pm$0.008 & -7.34$\pm$2.53 & -1.59$\pm$2.51 & RRab & 0 \\  
uks1     & 617437 & -- & 13059056015334 & 17:54:24.00 & -24:08:18.0 & 0.84 & 0.599870 & 0.379 &       --      & 18.710$\pm$0.088 & 17.726$\pm$0.021 & 16.709$\pm$0.030 & 16.096$\pm$0.008 &  1.37$\pm$2.10 & -1.09$\pm$2.24 & RRab & 0 \\  
uks1     & 617439 & -- & 13063152010908 & 17:54:26.24 & -24:10:35.0 & 1.88 & 0.651072 & 0.394 &       --      & 18.665$\pm$0.142 & 18.021$\pm$0.031 & 16.836$\pm$0.024 & 16.157$\pm$0.006 & -1.16$\pm$1.70 & -4.35$\pm$1.67 & RRab & 0 \\  
uks1     & 617456 & -- & 13054960007190 & 17:54:21.56 & -24:07:40.3 & 1.66 & 0.585990 & 0.307 &       --      &       --         & 17.924$\pm$0.019 & 16.730$\pm$0.021 & 16.117$\pm$0.006 & -2.16$\pm$1.76 & -3.73$\pm$1.75 & RRab & 0 \\  
uks1     & 617458 & -- & 13054960001006 & 17:54:23.20 & -24:07:34.6 & 1.46 & 0.574508 & 0.384 &       --      &       --         & 17.913$\pm$0.019 & 16.765$\pm$0.020 & 16.257$\pm$0.007 & -2.73$\pm$2.09 & -4.96$\pm$1.98 & RRab & 0 \\  
uks1     & 617866 & -- & 13063153006904 & 17:54:43.10 & -24:10:16.7 & 3.95 & 0.562057 & 0.337 &       --      &       --         & 18.580$\pm$0.019 & 17.183$\pm$0.022 & 16.386$\pm$0.007 & -1.96$\pm$1.83 & -2.62$\pm$1.71 & RRab & 0 \\  
    \hline
vvvcl160 & 1055884 & -- & 12641299011387 & 18:06:54.74 & -20:01:46.8 & 1.23 & 0.669990 & 0.240 & 18.930$\pm$0.044 & 17.429$\pm$0.023 & 16.031$\pm$0.005 & 14.860$\pm$0.009 & 14.226$\pm$0.002 & -3.41$\pm$0.52 &-15.90$\pm$0.54 & RRab & 1 \\  
vvvcl160 & 1055888 & -- & 12641299001075 & 18:06:56.16 & -20:00:39.3 & 0.20 & 0.563220 & 0.226 & 19.893$\pm$0.127 & 18.276$\pm$0.031 & 16.932$\pm$0.014 & 15.498$\pm$0.014 & 14.794$\pm$0.003 & -3.01$\pm$1.06 &-15.44$\pm$0.95 & RRab & 0 \\  
vvvcl160 & 1055898 & -- & 12641299008810 & 18:06:57.11 & -20:01:09.4 & 0.49 & 0.789131 & 0.162 & 19.732$\pm$0.050 & 18.074$\pm$0.021 & 16.548$\pm$0.006 & 15.196$\pm$0.009 & 14.500$\pm$0.002 & -2.42$\pm$0.62 &-16.87$\pm$0.58 & RRab & 0 \\  
vvvcl160 & 1055913 & -- & 12637204003016 & 18:07:00.81 & -19:58:53.0 & 2.00 & 0.558829 & 0.255 & 19.661$\pm$0.071 & 18.389$\pm$0.027 & 16.794$\pm$0.008 & 15.471$\pm$0.010 & 14.831$\pm$0.003 & -2.09$\pm$0.86 &-18.04$\pm$0.80 & RRab & 0 \\  
    \hline
  \end{tabular}
  } \tablefoot{This table is available in its entirety in electronic
    form at the CDS. A portion is shown here for guidance regarding
    its form and content.\\ \tablefoottext{a}{Flag 0 marks those
      RR~Lyrae stars consider cluster bonafide members. Flag 1 marks
      those RR~Lyrae stars whose membership is dubious. Flag 2 marks
      those RR~Lyrae stars which appear in the CVSGC, but our analysis
      considers as non-members. Flag 3 marks those RR~Lyrae stars with
      mismatch type between VIVACE and OGLE/CVSGC.}  }
\end{table*}

\begin{figure*}
  \section{Color excesses and distances to the VVV GCs}
  \centering
  \includegraphics[scale=1.0]{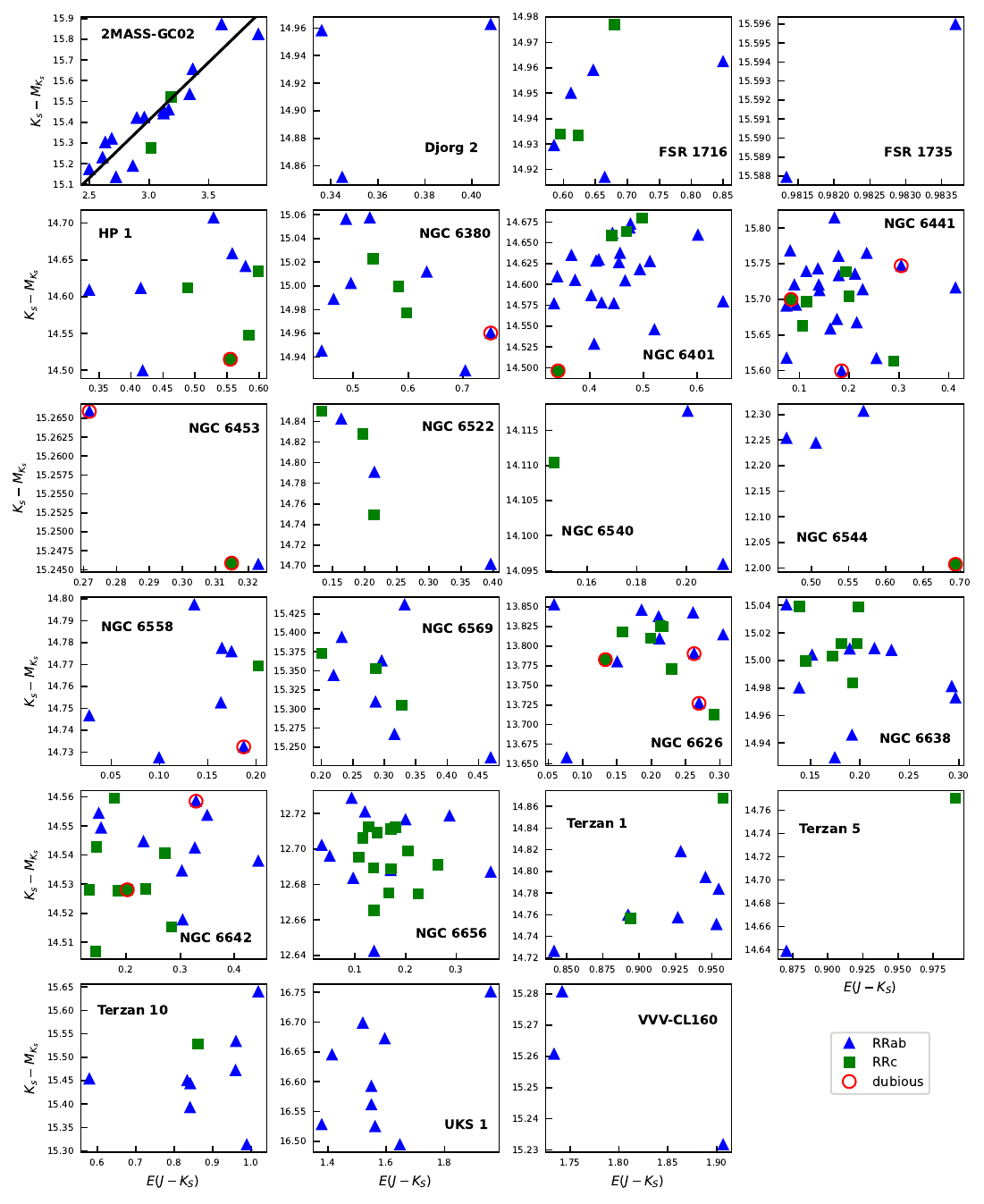}
  \caption{Apparent distance moduli versus color excesses for the RR~Lyrae cluster members based on their near-infrared PLZ relations. RRab stars are shown as blue triangles, and RRc stars as green squares. Red circles enclose dubious candidates, which are not plotted if they exhibit large offsets. The significant differential reddening in 2MASS-GC02 leads to the high dispersion observed among its RR~Lyrae members in the first panel. A solid black line shows the linear fit used to define the distance modulus and the selective-to-total extinction ratio to this GC (see Sect.~\ref{sec_distance2})}
  \label{fig_dis1}
\end{figure*}

\begin{table*}
  \caption{Color excesses and distances to the VVV GCs.}
  \label{tab_gcdis}
  \centering
  \resizebox{\textwidth}{!}{
  \begin{tabular}{ccccccccc}
    \hline\hline
    Cluster & $E(Z-K_s)$ & $E(Y-K_s)$ & $E(J-K_s)$ & $E(H-K_s)$ & $R_{\odot}$ & $R_{GC}$ & $R_{\odot,\rm GGCD}$ & $R_{GC,\rm GGCD}$\\
     & (mag) & (mag) & (mag) & (mag) & (kpc) & (kpc) & (kpc) & (kpc)\\
    \hline
    2MASS-GC02 & -- & 3.73$\ \pm\ $0.21 & 3.04$\ \pm\ $0.37 & 1.08$\ \pm\ $0.15 & 6.3$\ \pm\ $0.4 & 4.25$\ \pm\ $0.07 & 5.50$\ \pm\ $0.44 & 2.91$\ \pm\ $0.37 \\
    BH~261 & 0.18$\ \pm\ $0.01 & 0.03$\ \pm\ $0.01 & 0.18$\ \pm\ $0.01 & 0.01$\ \pm\ $0.01 & 7.05$\ \pm\ $0.18 & 3.02$\ \pm\ $0.05 & 6.12$\ \pm\ $0.26 & 2.20$\ \pm\ $0.23 \\
    Djorg~2 & 0.81$\ \pm\ $0.09 & 0.47$\ \pm\ $0.09 & 0.36$\ \pm\ $0.03 & 0.09$\ \pm\ $0.02 & 9.09$\ \pm\ $0.20 & 1.03$\ \pm\ $0.17 & 8.76$\ \pm\ $0.18 & 0.80$\ \pm\ $0.13 \\
    FSR~1716 & 1.86$\ \pm\ $0.13 & 1.19$\ \pm\ $0.10 & 0.66$\ \pm\ $0.08 & 0.18$\ \pm\ $0.09 & 8.52$\ \pm\ $0.10 & 0.54$\ \pm\ $0.05 & 7.43$\ \pm\ $0.27 & 4.14$\ \pm\ $0.02 \\
    FSR~1735 & 2.74$\ \pm\ $0.12 & 1.77$\ \pm\ $0.09 & 0.98$\ \pm\ $0.01 & 0.22$\ \pm\ $0.03 & 10.72$\ \pm\ $0.03 & 2.59$\ \pm\ $0.03 & 9.08$\ \pm\ $0.53 & 3.25$\ \pm\ $0.22 \\
    HP~1 & 1.46$\ \pm\ $0.07 & 0.94$\ \pm\ $0.08 & 0.50$\ \pm\ $0.09 & 0.14$\ \pm\ $0.04 & 7.57$\ \pm\ $0.18 & 1.52$\ \pm\ $0.06 & 7.00$\ \pm\ $0.14 & 1.26$\ \pm\ $0.13 \\
    NGC~6380 & 1.51$\ \pm\ $0.05 & 0.93$\ \pm\ $0.08 & 0.55$\ \pm\ $0.08 & 0.15$\ \pm\ $0.06 & 8.99$\ \pm\ $0.20 & 0.83$\ \pm\ $0.18 & 9.61$\ \pm\ $0.30 & 2.15$\ \pm\ $0.21 \\
    NGC~6401 & 1.13$\ \pm\ $0.10 & 0.73$\ \pm\ $0.08 & 0.45$\ \pm\ $0.07 & 0.13$\ \pm\ $0.03 & 7.72$\ \pm\ $0.15 & 0.79$\ \pm\ $0.10 & 7.44$\ \pm\ $0.22 & 1.03$\ \pm\ $0.14 \\
    NGC~6441 & 0.46$\ \pm\ $0.15 & 0.23$\ \pm\ $0.13 & 0.16$\ \pm\ $0.06 & 0.02$\ \pm\ $0.03 & 13.3$\ \pm\ $0.5 & 5.2$\ \pm\ $0.5 & 12.73$\ \pm\ $0.16 & 4.78$\ \pm\ $0.15 \\
    NGC~6453 & 0.82$\ \pm\ $0.01 & 0.53$\ \pm\ $0.01 & 0.32$\ \pm\ $0.01 & 0.12$\ \pm\ $0.01 & 10.49$\ \pm\ $0.08 & 2.36$\ \pm\ $0.07 & 10.07$\ \pm\ $0.22 & 2.10$\ \pm\ $0.19 \\
    NGC~6522 & 0.58$\ \pm\ $0.20 & 0.36$\ \pm\ $0.14 & 0.22$\ \pm\ $0.08 & 0.06$\ \pm\ $0.06 & 8.7$\ \pm\ $0.3 & 1.26$\ \pm\ $0.16 & 7.29$\ \pm\ $0.21 & 1.04$\ \pm\ $0.17 \\
    NGC~6540 & 0.41$\ \pm\ $0.12 & 0.17$\ \pm\ $0.09 & 0.19$\ \pm\ $0.03 & 0.00$\ \pm\ $0.01 & 6.53$\ \pm\ $0.03 & 1.89$\ \pm\ $0.03 & 5.91$\ \pm\ $0.27 & 2.34$\ \pm\ $0.25 \\
    NGC~6544 & 1.05$\ \pm\ $0.34 & 0.63$\ \pm\ $0.20 & 0.51$\ \pm\ $0.04 & 0.29$\ \pm\ $0.10 & 2.59$\ \pm\ $0.03 & 5.70$\ \pm\ $0.03 & 2.58$\ \pm\ $0.06 & 5.62$\ \pm\ $0.06 \\
    NGC~6558 & 0.32$\ \pm\ $0.13 & 0.13$\ \pm\ $0.14 & 0.14$\ \pm\ $0.05 & 0.02$\ \pm\ $0.02 & 8.67$\ \pm\ $0.16 & 0.73$\ \pm\ $0.10 & 7.79$\ \pm\ $0.18 & 0.93$\ \pm\ $0.06 \\
    NGC~6569 & 0.63$\ \pm\ $0.07 & 0.36$\ \pm\ $0.07 & 0.30$\ \pm\ $0.07 & 0.10$\ \pm\ $0.06 & 10.7$\ \pm\ $0.4 & 2.6$\ \pm\ $0.3 & 10.53$\ \pm\ $0.26 & 2.59$\ \pm\ $0.23 \\
    NGC~6626 & 0.47$\ \pm\ $0.10 & 0.29$\ \pm\ $0.08 & 0.20$\ \pm\ $0.07 & 0.03$\ \pm\ $0.04 & 5.43$\ \pm\ $0.19 & 2.90$\ \pm\ $0.18 & 5.37$\ \pm\ $0.10 & 3.02$\ \pm\ $0.09 \\
    NGC~6638 & 0.48$\ \pm\ $0.10 & 0.33$\ \pm\ $0.09 & 0.19$\ \pm\ $0.05 & 0.03$\ \pm\ $0.03 & 9.40$\ \pm\ $0.21 & 1.31$\ \pm\ $0.19 & 9.78$\ \pm\ $0.34 & 2.30$\ \pm\ $0.25 \\
    NGC~6642 & 0.40$\ \pm\ $0.14 & 0.25$\ \pm\ $0.11 & 0.24$\ \pm\ $0.09 & 0.06$\ \pm\ $0.07 & 7.62$\ \pm\ $0.19 & 0.92$\ \pm\ $0.13 & 8.05$\ \pm\ $0.20 & 1.66$\ \pm\ $0.01 \\
    NGC~6656 & 0.34$\ \pm\ $0.09 & 0.23$\ \pm\ $0.08 & 0.15$\ \pm\ $0.06 & 0.05$\ \pm\ $0.08 & 3.30$\ \pm\ $0.05 & 5.06$\ \pm\ $0.05 & 3.30$\ \pm\ $0.04 & 5.00$\ \pm\ $0.03 \\
    Terzan~1 & 2.43$\ \pm\ $0.12 & 1.62$\ \pm\ $0.10 & 0.92$\ \pm\ $0.04 & 0.31$\ \pm\ $0.05 & 7.52$\ \pm\ $0.13 & 1.13$\ \pm\ $0.8 & 5.67$\ \pm\ $0.17 & 2.52$\ \pm\ $0.17 \\
    Terzan~4 & 2.87$\ \pm\ $0.03 & 1.95$\ \pm\ $0.03 & 1.14$\ \pm\ $0.01 & 0.39$\ \pm\ $0.02 & 9.43$\ \pm\ $0.13 & 1.47$\ \pm\ $0.10 & 7.59$\ \pm\ $0.31 & 0.82$\ \pm\ $0.02 \\
    Terzan~5 & 2.53$\ \pm\ $0.22 & 1.59$\ \pm\ $0.17 & 0.93$\ \pm\ $0.06 & 0.35$\ \pm\ $0.01 & 7.26$\ \pm\ $0.11 & 1.55$\ \pm\ $0.06 & 6.62$\ \pm\ $0.15 & 1.65$\ \pm\ $0.13 \\
    Terzan~9 & 2.19$\ \pm\ $0.01 & 1.45$\ \pm\ $0.01 & 0.84$\ \pm\ $0.01 & 0.27$\ \pm\ $0.01 & 7.36$\ \pm\ $0.05 & 1.25$\ \pm\ $0.03 & 5.77$\ \pm\ $0.34 & 2.46$\ \pm\ $0.33 \\
    Terzan~10 & 2.46$\ \pm\ $0.35 & 1.52$\ \pm\ $0.21 & 0.88$\ \pm\ $0.12 & 0.30$\ \pm\ $0.09 & 10.4$\ \pm\ $0.4 & 2.5$\ \pm\ $0.4 & 10.21$\ \pm\ $0.40 & 2.17$\ \pm\ $0.37 \\
    UKS~1 & -- & 2.39$\ \pm\ $0.18 & 1.57$\ \pm\ $0.16 & 0.56$\ \pm\ $0.09 & 15.4$\ \pm\ $0.7 & 7.4$\ \pm\ $0.7 & 15.58$\ \pm\ $0.56 & 7.47$\ \pm\ $0.54 \\
    VVV-CL~160 & 4.56$\ \pm\ $0.12 & 3.13$\ \pm\ $0.05 & 1.79$\ \pm\ $0.08 & 0.62$\ \pm\ $0.03 & 7.95$\ \pm\ $0.12 & 1.79$\ \pm\ $0.03 & 6.80$\ \pm\ $0.50 & 1.91$\ \pm\ $0.31 \\
    \hline
  \end{tabular}
  }
\end{table*}
\end{appendix}

\end{document}